\documentclass[fleqn,usenatbib]{mnras}

\usepackage{newtxtext,newtxmath}

\usepackage[T1]{fontenc}

\DeclareRobustCommand{\VAN}[3]{#2}
\let\VANthebibliography\thebibliography
\def\thebibliography{\DeclareRobustCommand{\VAN}[3]{##3}\VANthebibliography}

\usepackage{graphicx}	
\usepackage{amsmath}	
\usepackage{algorithm}
\usepackage{algpseudocode}
\usepackage{booktabs}
\usepackage{natbib}

\title[Artificial Intelligence for AGN Recognition]{An Active Galactic Nucleus Recognition Model based on Deep Neural Network}

\author[Bo-Han Chen et al.]{
Bo Han Chen,$^{1}$\thanks{E-mail:369grant2@gmail.com}
Tomotsugu Goto$^{1,2}$,
Seong Jin Kim$^{1,2}$,
Ting Wen Wang$^{2}$,
\and
Daryl Joe D. Santos$^{2}$,
Simon C.-C. Ho$^{2}$,
Tetsuya Hashimoto$^{1,3}$,
Artem Poliszczuk$^{4}$,
\and
Agnieszka Pollo$^{4,5}$,
Sascha Trippe$^{6}$,
Takamitsu Miyaji$^{7,8}$\footnote{On sabbatical leave from IA-UNAM-E at AIP.},
Yoshiki Toba$^{9,10,11}$,
\and
Matthew Malkan$^{12}$,
Stephen Serjeant$^{13}$,
Chris Pearson$^{14,15,16}$,
Ho Seong Hwang$^{17}$,
\and
Eunbin Kim$^{17}$,
Hyunjin Shim$^{18}$,
Ting-Yi Lu$^{2}$,
Tiger Y.-Y. Hsiao$^{2}$,
Ting-Chi Huang$^{19,20}$,
\and
Martín Herrera-Endoqui$^{7}$,
Blanca Bravo-Navarro$^{7,21}$
and Hideo Matsuhara$^{19,20}$
\\
$^{1}$Department of Physics, National Tsing Hua University, No. 101, Section 2, Kuang-Fu Road, Hsinchu City 30013, Taiwan\\
$^{2}$Institute of Astronomy, National Tsing Hua University, No. 101, Section 2, Kuang-Fu Road, Hsinchu City 30013, Taiwan\\
$^{3}$Centre for Informatics and Computation in Astronomy (CICA), National Tsing Hua University, 101, Section 2. Kuang-Fu Road, Hsinchu, 30013, Taiwan\\
$^{4}$National Centre for Nuclear Research, ul.Pasteura 7, 02-093 Warsaw, Poland\\
$^{5}$Astronomical Observatory of the Jagiellonian University, ul.Orla 171, 30-244 Krakow, Poland\\
$^{6}$Department of Physics and Astronomy, Seoul National University, 1, Gwanak Road, Seoul, 08826, Republic of Korea\\
$^{7}$Instituto de Astrnom\'ia sede Ensenada, Universidad Nacinal Aut\'onoma de M\'exico (IA-UNAM-E) Km 107, Carret. Tij.-Ens., 22860, Ensenada,BC, Mexico\\
$^{8}$Leibnitz Instituto f\"ur Astrophysik (AIP), An der Sternwarte 16, 14482, Potsdam, Germany \\
$^{9}$Department of Astronomy, Kyoto University, Kitashirakawa-Oiwake-cho, Sakyo-ku, Kyoto 606-8502, Japan\\
$^{10}$Academia Sinica Institute of Astronomy and Astrophysics, 11F of Astronomy-Mathematics Building, AS/NTU, No.1, Section 4, Roosevelt Road, Taipei 10617, Taiwan\\
$^{11}$Research Center for Space and Cosmic Evolution, Ehime University, 2-5 Bunkyo-cho, Matsuyama, Ehime 790-8577, Japan\\
$^{12}$Department of Physics and Astronomy, UCLA, 475 Portola Plaza, Los Angeles, CA 90095-1547, USA\\
$^{13}$School of Physical Sciences, The Open University, Milton Keynes, MK7 6AA, UK\\
$^{14}$RAL Space, STFC Rutherford Appleton Laboratory, Didcot, Oxon, OX11 0QX, UK\\
$^{15}$The Open University, Milton Keynes, MK7 6AA, UK\\
$^{16}$University of Oxford, Keble Rd, Oxford, OX1 3RH, UK\\
$^{17}$Korea Astronomy and Space Science Institute, 776 Daedeokdae-ro, Yuseong-gu, Daejeon 34055, Republic of Korea\\
$^{18}$Department of Earth Science Education, Kyungpook National University, 80 Daehak-ro, Buk-gu, Daegu 41566, Republic of Korea\\
$^{19}$Department of Space and Astronautical Science, Graduate University for Advanced Studies, SOKENDAI, Shonankokusaimura, Hayama, Miura\\ District, Kanagawa 240-0193, Japan\\
$^{20}$Institute of Space and Astronautical Science, Japan Aerospace Exploration Agency, 3-1-1 Yoshinodai, Chuo-ku, Sagamihara, Kanagawa 252-5210, Japan\\
$^{21}$Inginiero Aeroespacial, 
Universidad Aut\'onoma de Baja California, Blvd. Universitario 1000 Valle de Las Palmas, Tijuana, B.C. 22260, Mexico\\
}

\date{Accepted 2020 December 10. Received 2020 December 9; in original form 2020 September 14}

\pubyear{2020}

\begin{document}
\label{firstpage}
\pagerange{\pageref{firstpage}--\pageref{lastpage}}
\maketitle
\begin{abstract}
To understand the cosmic accretion history of supermassive black holes, separating the radiation from active galactic nuclei (AGNs) and star-forming galaxies (SFGs) is critical.  However, a reliable solution on photometrically recognising AGNs still remains unsolved.  In this work, we present a novel AGN recognition method based on Deep Neural Network (Neural Net; NN).  The main goals of this work are (i) to test if the AGN recognition problem in the North Ecliptic Pole Wide (NEPW) field could be solved by NN; (ii) to shows that NN exhibits an improvement in the performance compared with the traditional, standard spectral energy distribution (SED) fitting method in our testing samples; and (iii) to publicly release a reliable AGN/SFG catalogue to the astronomical community using the best available NEPW data, and propose a
better method that helps future researchers plan an advanced NEPW database.  Finally, according to our experimental result, the NN recognition accuracy  is around 80.29\% - 85.15\%, with AGN completeness around 85.42\% - 88.53\% and SFG completeness around  81.17\% - 85.09\%.
\end{abstract}

\begin{keywords}
galaxies: active
surveys
methods: data analysis
ultraviolet: galaxies
infrared: galaxies
submillimetre: galaxies
\end{keywords}



\section{Introduction}

An active galactic nucleus (AGN) is a compact region at the centre of a galaxy which is highly-luminous due to processes not caused by star-forming activities. It is widely believed that AGNs are powered by the accretion of super massive black holes (SMBHs) located at the centre of galaxies.  Furthermore, it is found that the bulge masses of galaxies co-evolve with the mass of the black holes (e.g. \citealt{Magorrian1998}).  Thus, studying AGNs can help us understand galaxy evolution.

In order to reveal the cosmic accretion history of SMBHs, it is crucial to find AGNs in the universe. However, it has been notoriously difficult to identify AGNs from normal SFGs photometrically. The difficulty comes from two aspects. First, UV and X-ray observations usually suffer from the extinction by dust and the absorption by gas surrounding AGNs. (e.g. \citealt{Webster1995}; \citealt{Alexander2001}; \citealt{Richards2003}).  Though the extinction-free observations in mid-infrared (MIR) bands are promising alternative, MIR includes both polycyclic aromatic hydrocarbon (PAH) emissions from SFGs and power-law emission from AGNs.  Thus, a definite classification based on MIR data could only be performed by using spectroscopic data but not photometric data, while the former is usually not available.   Therefore, finding a way to separate AGNs from SFGs photometrically is important to advance the field.

There are several photometric and spectroscopic methods proposed to select AGNs. Regarding photometric methods, one of them is using MIR colours from the Spitzer-WISE Survey (\citealt{Lacy2004}; \citealt{Stern2005}; \citealt{Richards2006}) or optical colours from Baryon Oscillation Spectroscopic Survey (BOSS) (\citealt{Ross2012}).  Another is the variability selection based on \emph{ugriz} optical bands in the Sloan Digital Sky Survey (SDSS) region (\citealt{PalanqueDelabrouille2011}).  The other is via spectral energy distribution (SED) fitting, which covers the mid-IR wavelength gap and includes up to 36 band filters using the AKARI space telescope(\citealt{Huang2017}, \citet{Wang2020}).  In addition, a different study used fuzzy support vector machine (FSVM), which is a machine learning-based method, and it provided a high quality result on North Ecliptic Pole Deep (NEPD) field using 8 filters including 3 NIR bands and 5 MIR bands of the AKARI. (\citealt{Poliszczuk2019}).  In terms of spectroscopic AGN selection methods, some selections of local AGNs are done by using BPT diagnostic(\citealt{Baldwin1981}; \citealt{Veilleux1987}). For selections of high redshift AGNs, \citet{Yan2011} select AGNs by combining the [O{\scriptsize III}]/H$\beta$ ratio with rest-frame $U - B$ color. \citet{Juneau2011} and \citet{Juneau2013} developed mass-excitation diagnostic to select AGNs with redshift > 0.3. \citet{Marocco2011} selected AGNs from the SDSS by using spectral classification.  Finally, \citet{Zhang2018} proposed a kinematics–excitation (KEx) diagram to select AGNs. \citet{Zhang2019} select AGNs at intermediate redshift (z=0.3–0.8) by using supervised machine learning classification algorithms.
\begin{figure}
	\includegraphics[width=\columnwidth]{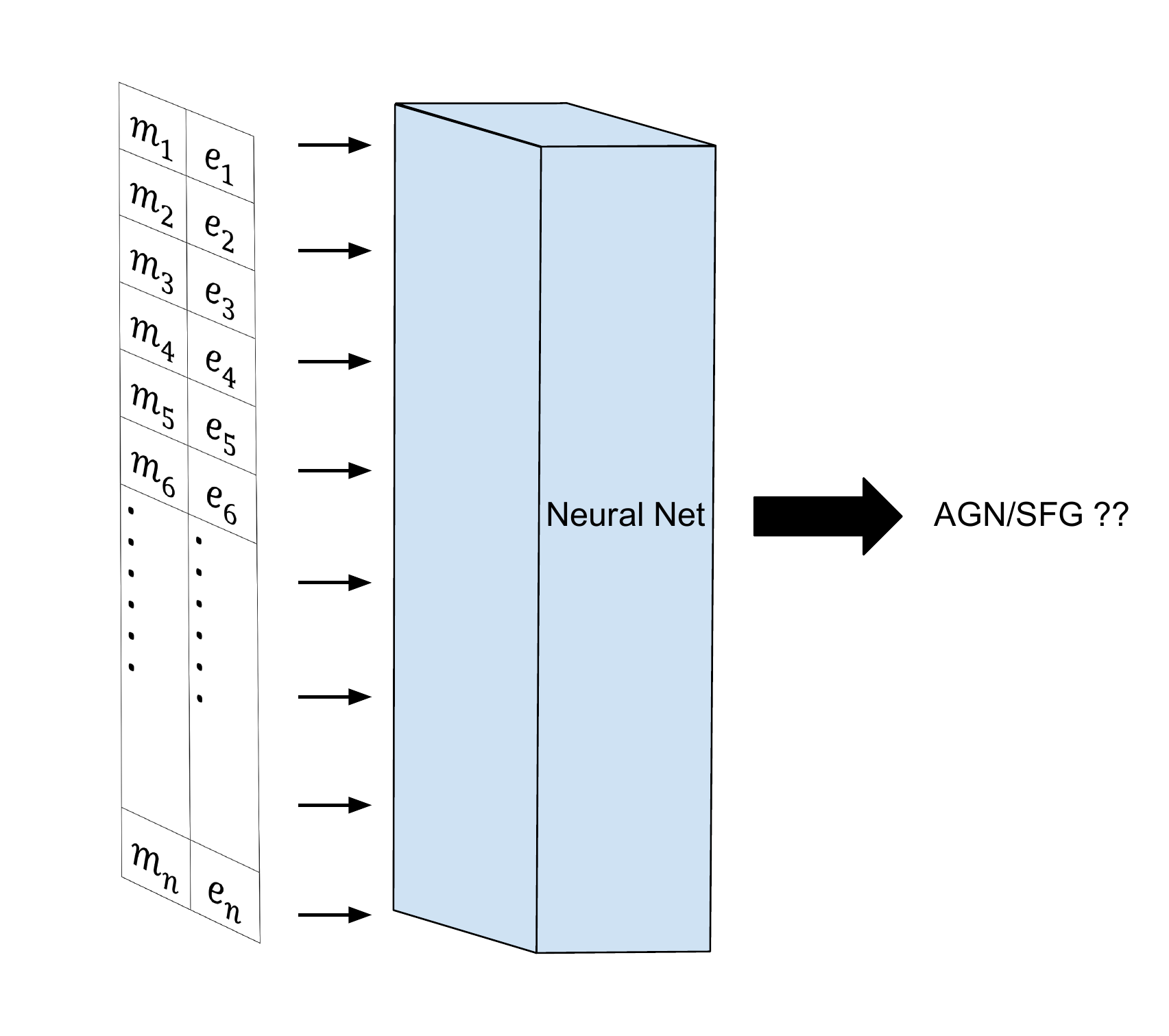}
    \caption{The NN takes several photometric magnitudes ($m_1$,$m_2$...) and errors  ($e_1$,$e_2$...) data of a galaxy as input, and accurately states whether the inputted galaxy is an AGN or a SFG.}
    \label{fig:figure1}
\end{figure}

In this paper, we introduce a state-of-the-art technique, Deep Neural Network (Neural Net; NN), to build a robust model that can recognise AGNs from star-forming galaxies (SFGs).  NN is a kind of algorithms inspired by biological neural networks that constitute animal brains. NN imitates biological neural network connections by proceeding linear matrix operation and biological neuron by applying a specific non-linear function.  We describe the details of our NN in Section~\ref{sec:Model Architecture}.  Our goal is to construct a NN that can take several photometric magnitudes and errors of a galaxy as an input, and accurately state if the galaxy is an AGN or a SFG (Fig.~\ref{fig:figure1}). It is widely known that NN is good at solving a specific problem, such as image classification (\citealt{Krizhevsky2017}),  young stellar objects search (\citealt{Chiu2020}), or even redshift estimation (\citealt{Collister2004}, \citealt{DeWei2019}).

A sufficiently large training set which includes the input data and the corresponding true answer (ground truth) is necessary to train the NN algorithm.  In our work, the input data consists of at most 44 band magnitudes and errors, which includes observations from Hyper Suprime-Cam (HSC), $AKARI$, Maidanak, Canada-France-Hawaii Telescope (CFHT), Kitt Peak National Observatory (KPNO), Wide-field Infrared Survey Explorer ($WISE$), $Spitzer$, and $Herschel$ (\citealt{Kim2020}). The ground truth is taken from X-ray (\citealt{Krumpe2014}) and spectroscopic (\citealt{Shim2013}) classifications. We describe the details of the data in Section~\ref{sec:Sample selection}.  There are 1,870 galaxy classification ground truths in total; about 10\% of the galaxies are assigned as validation samples, which means they do not participate in training and are only applied for validating the accuracy of the model.

Above all, the points of this work are as follows.

\begin{itemize}
\item NN could be applied for solving the AGN recognition problem in the NEPW field.
\item We verify that the proposed NN method is superior to the popular SED fitting methodology in the testing samples from the NEPW field.
\item We publicly release a more reliable AGN/SFG catalogue using the best available NEPW data.
\end{itemize}

It is known through the universal approximation theorem that NN can approximate any given multivariate polynomial and any generic non-linearity (\citealt{Cybenko1989}, \citealt{Hornik1991}, \citealt{Lu2017}, see also  \citealt{Lin2017a}) therefore NN is expected to be able to perform well in photometric classification problems in general.  In addition, the performance of NN would be sustainingly reinforced as the number of training data increasing (\citealt{Ng2017}).  Hence, with the expected development of the training sample number and the upcoming observation in the NEPW field in near future (e.g. eROSITA, Subaru/PFS...), we could look ahead to a steady advancement on this project based on our method.   Our aim in this paper is not to compare with other machine learning model against NN and show that NN is the most efficient one at the current stage, but rather to test whether NN can be used in selecting AGN.  Once we verify that NN can be also used for our NEPW data and performs better than traditional SED fitting method, it could help the community invest more resources on developing the size of the training set, consequently leading to a steady development of the AGN recognition  project.

This work is organised as follows. We describe our sample selection and NN model in Section~\ref{sec:Data And Model Structure}. Our AGN recognition results are described in Section~\ref{sec:Empirical Result}. We present the discussion in Section~\ref{sec:discussion}. Our conclusions are in Section~\ref{sec:Conclusions}. Throughout this paper, we use AB magnitude system unless otherwise mentioned.

\section{Data And Model Structure}
\label{sec:Data And Model Structure}

\subsection{Sample selection}
\label{sec:Sample selection} 

All involving galaxy samples in this work are based on a multi-wavelength catalogue  in the NEPW field (\citealt{Kim2020}).   The catalogue consists of various photometric data from optical CFHT/$u$-band to the Hershel/SPIRE bands, obtained to support the AKARI NEPW survey ($5.4$ deg$^{2}$) data, centred at  ($\rm{RA}=18h00m00s, \rm{Dec}.=+66^{\circ}33'38''$;   \citealt{Matsuhara2006}; \citealt{Lee2009};  \citealt{Kim2012}). 

\begin{figure}
	\includegraphics[width=1.0\columnwidth]{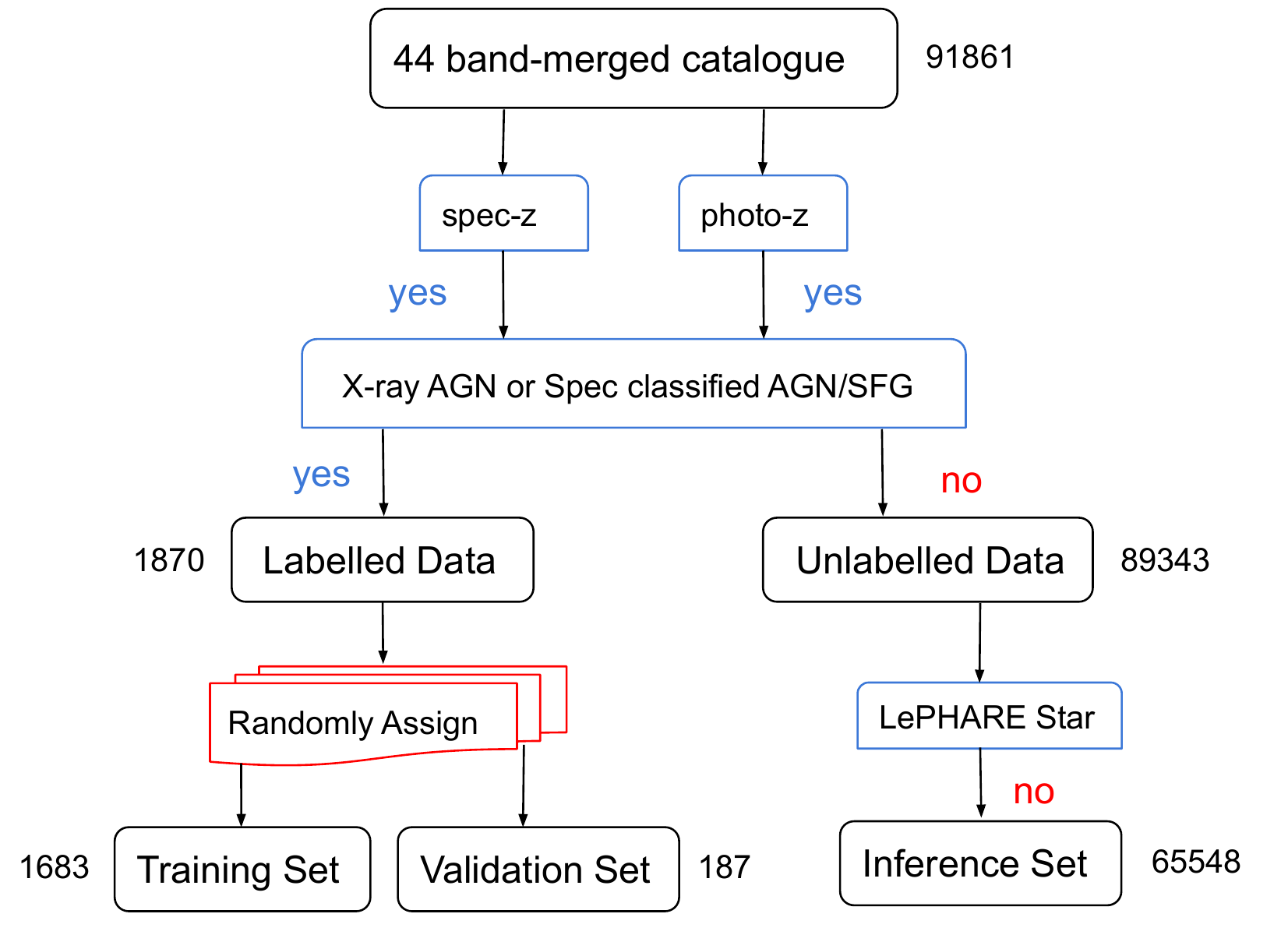}
    \caption{The flow chart of data preprocessing. The "Labelled Data" are randomly and equally divided into 10 groups; whenever the training is performed, one of the groups (i.e., 10\% of data) is excluded from the training to serve as validation data.  The inference set contains galaxy data that will be classified with our trained NN.}  
    \label{fig:figure2}
\end{figure}

The procedure for data preprocessing is shown in Fig.~\ref{fig:figure2}.  The catalogue contains 91,861 sources in total, and 2,026 of them have spectroscopic data.  The spectroscopic data is provided by \citet{Miyaji2019}, \citet{Oi2017} and \citet{Shim2013}.  

In our study, we excluded objects which have neither spectroscopic nor photometric redshift measurements.  The photometric redshifts of our samples without spectroscopic redshifts are estimated using $LePhare$ (\citealt{Ho2020}), a set of FORTRAN commands to compute photometric redshifts and to perform SED fitting.

Among the sources with spectroscopic data in the multi-wavelength catalogue (\citealt{Kim2020}), 1615 SFGs and 255 AGNs are already classified.  The identification comes from two sources.  The first one is the analysis of spectroscopic data, obtained by MMT/Hectospec and WIYN/Hydra.  The observed spectra were classified via visual inspection and/or identification of the diagnostics with emission lines (\citealt{Shim2013}).  The second source is the analysis of X-ray data.  By cross-matching X-ray sources from $Chandra$ North Ecliptic Pole Deep (NEPD) survey counterpart and the MIR-selected AGN candidates counterpart from AKARI NEPW field survey, a set of objects are confirmed as AGNs if X-ray sources have X-ray luminosity of $L_x > 10^{41.5} erg\ s^{-1}$ in a standard X-ray band (e.g. 2-10 keV or  0.5-2 keV) (\citealt{Krumpe2014}).  84\% of the AGN samples are provided spectroscopically, and X-ray identify 30\% of the AGN samples.  Roughly 14\% of AGNs are consistently identified by the two methods.  Total number of $1615 + 255 = 1870$ objects provide us with a foothold to train our model by supervised learning. We denote these identified objects as "Labelled Data"; on the other hand, the unidentified objects are denoted as "Unlabelled Data".   

We use $LePHARE$ classification to remove stars in the "Unlabelled Data".
The SED of stellar templates and galaxy templates are used here.  When SED fitting is performed (\citealt{Ho2020}), $\chi^2$ value is evaluated for both the galaxy templates (\citealt{Ilbert2008}) and stellar templates (\citealt{Bohlin1995}, \citealt{Pickles1998}, \citealt{Chabrier2000}) for each source. Then, they compare the two $\chi^2$ values to separate stars and galaxies. If $\chi^2_{gal} > \chi^2_{star}$, where $\chi^2_{gal}$ and $\chi^2_{star}$ are the minimum $\chi^2$ values obtained with the galaxy and stellar templates, respectively, the object is flagged as a star.  Here 23795 stars are removed, and the remaining 65548 galaxy objects would be classified as either AGN or SFG in Sec.~\ref{sec:The inference result on whole NEP field}

\begin{figure*}
	\includegraphics[width=0.8\textwidth]{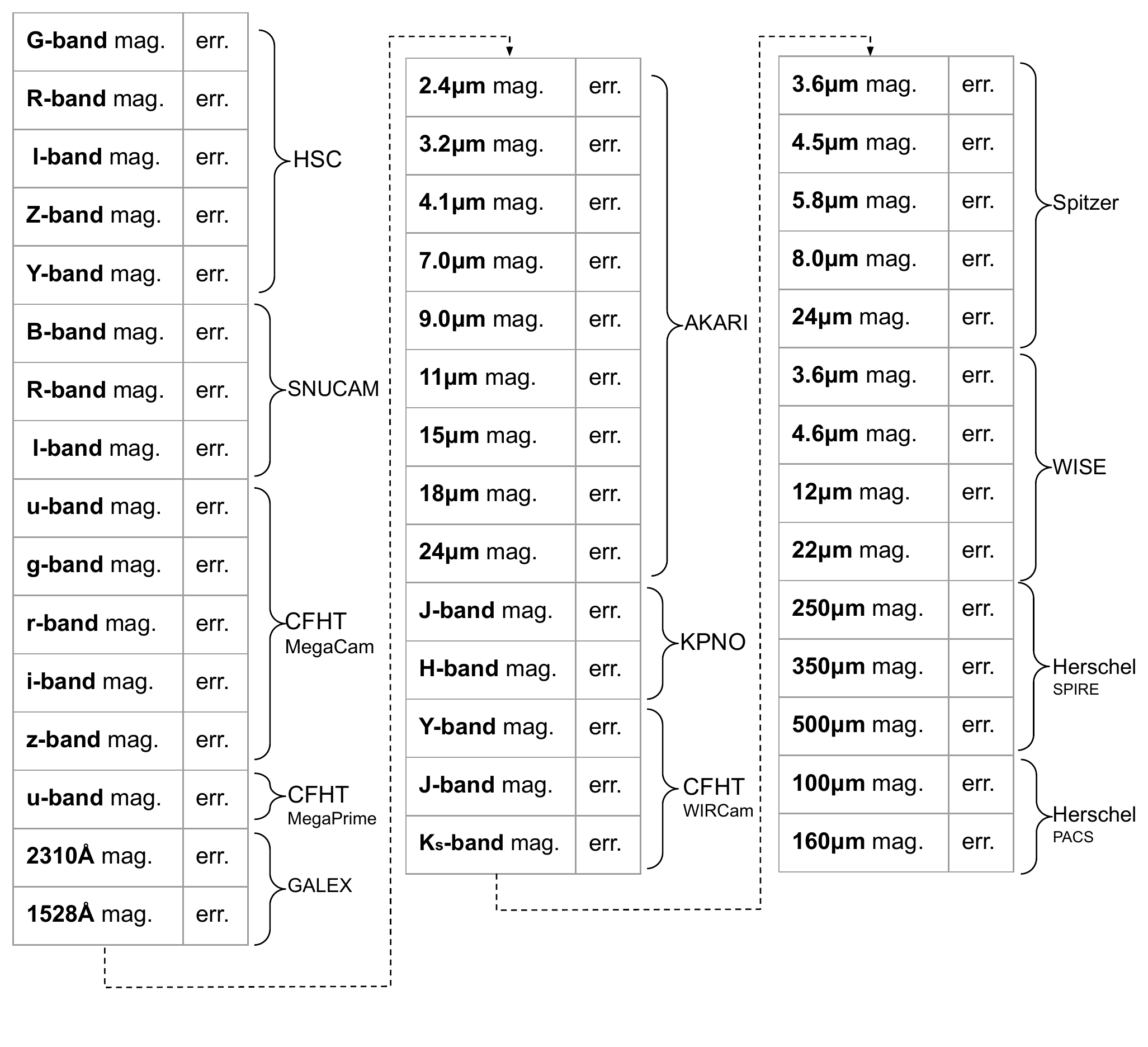}
    \caption{The schematic diagram of the inputted band data.  Each band has a magnitude and an error.  The whole diagram form a $(44 \times 2)$ array.}
    \label{fig:figure3}
\end{figure*}

In terms of the input data of the NN, including those aimed for training, testing, or merely inferring, we use all available photometric bands in multi-wavelength catalogue. We provide a summary of the photometric bands used in this study in Fig.~\ref{fig:figure3}.   The observational details are described in the following subsections. In addition, a more detailed description can found in \citet{Kim2020}.

\subsubsection{The 44 Band-Merged data}
\label{sec:The 44 Band-Merged data} 

In our 44 band-merged catalogue, the data from UV to optical are provided by $Subaru$ Hyper Suprime-Cam (HSC), Maidanak’s Seoul National University 4K$\times$4K Camera (SNUCAM), Canada-France-Hawaii Telescope (CFHT) MegaCam, MegaPrime and Galaxy Evolution Explorer (GALEX).

Subaru is a telescope on the summit of Mauna Kea in Hawaii, and its HSC provides us at most five-band photomerties and their errors including \emph{g, r, i, z} and \emph{y} bands (\citealt{Oi2020}).  The photomerties have the detection limits at 28.6, 27.3, 26.7, 26.0 and 25.6, respectively.

SNUCAM is a charge-coupled device (CCD) camera located at the Maidanak observatory in Uzbekistan, providing \emph{B, R, I}-band magnitudes and errors in our input (\citealt{Jeon2010}).  The three detection limits respectively are 23.4, 23.1 and 22.3.

CFHT is a telescope also located atop the summit of Mauna Kea.  MegaCam and MegaPrime are optical imaging facilities at CFHT.  We use \emph{u, g, r, i, z}-band data from MegaCam (\citealt{Hwang2007};  \citealt{Oi2014}) and u-band data from MegaPrime (\citealt{Huang2020}).  Each of the detection limits from MegaCam are 26.0, 26.1, 25.6, 24.8 and 24.0.  The u-band detection limit from MegaPrime is 25.27.

GALEX is an UV space telescope providing the shortest wavelength data in our NN input.  It provides near-UV and far-UV band magnitudes and errors, respectively, corresponding to the wavelengths of 0.2310 and 0.1528 µm (\citealt{Martin2005}). The near-UV detection be of a limit 22.9, and the far-UV one is 22.2.

In Near-Infrared (NIR) to Mid-Infrared (MIR) data, we use the data obtained by $Spitzer$, Wide-field Infrared Survey Explorer ($WISE$), $AKARI$ Infrared Camera (IRC), Florida Multi-object Imaging Near-IR Grism Observational Spectrometer (FLAMINGOS) and CFHT WIRCam.

$Spitzer$ is an IR space telescope.  It provides us IRAC 1, IRAC 2, IRAC 3, IRAC 4, MIPS 24-band magnitudes and errors, which correspond to 3.6, 4.5, 5.8, 8.0 and 24 µm, respectively (\citealt{Nayyeri2018}).   The detection limit of the five bands are 21.8, 22.4, 16.6, 15.4 and 20.6, respectively.

$WISE$ is also an IR space telescope. Its observation includes W1, W2, W3, W4-band magnitudes and errors which correspond to wavelengths of 3.6, 4.6, 12 and 22 µm, respectively. (\citealt{Jarrett2011}).  Each of the detection limits from $WISE$ are 18.1, 17.2, 18.4 and 16.1.   Both $WISE$ and $Spitzer$ have a filter gap between 12 µm and 22 µm; in contrast, $AKARI$ provides us the data in this range.

$AKARI$ is another IR space telescope with the continuous wavelength coverage from NIR to MIR, thus provides us with the important information in recognising AGNs.  The IR camera of $AKARI$ includes \emph{N2, N3, N4, S7, S9W, S11, L15, L18W,} and \emph{L24}-band magnitudes and errors,which correspond to 2.4, 3.2, 4.1, 7, 9, 11, 15, 18 and 24 µm, respectively (\citealt{Kim2012}).   The detection carry out by $AKARI$ cameras have the corresponding limits equal to 20.9, 21.1, 21.1, 19.5, 19.3, 19.0, 18.6, 18.7 and 17.8.

The observations from FLAMINGOS, a wide-field IR imager and multi-slit spectrometer on the Kitt Peak National Observatory (KPNO), provide us \emph{J, H}-band magnitudes and errors (\citealt{Jeon2014}). The two detection limits respectively are 21.6 and 21.3.

CFHT also provides NIR data.  The data from CFHT WIRCam, a NIR mosaic imager, including \emph{Y, J, Ks}-band magnitudes and errors is used in this work (\citealt{Oi2014}).   The three observations from the imager correspondingly have the detection limits 23.4, 23.0 and 22.7.

In our data collection, the Far-Infrared (FIR) Data is uniquely provided by $Herschel$, a FIR and sub-millimetre space telescope.  Its two instruments, i.e., Spectral and Photometric Imaging Receiver (SPIRE) and Photodetector Array Camera and Spectrometer (PACS) are FIR imaging photometers.  The SPIRE has three bands, respectively centred on 250, 350 and 500 µm  and be of the detection limits 14, 14.2 and 13.8 .  In terms of PACS, two bands centred on 100 and 160 µm are included in this research, limiting at the magnitudes 14.7 and 14.1.  In summary, $Herschel$ provides us at most five-band photometries and their errors in FIR range (\citealt{Pearson2017}, \citealt{Pearson2018}).

\subsubsection{The statistical information regarding the Labelled data.}
\label{sec:The statistical information regarding the Labelled data.}

\begin{figure}
	\includegraphics[width=\columnwidth]{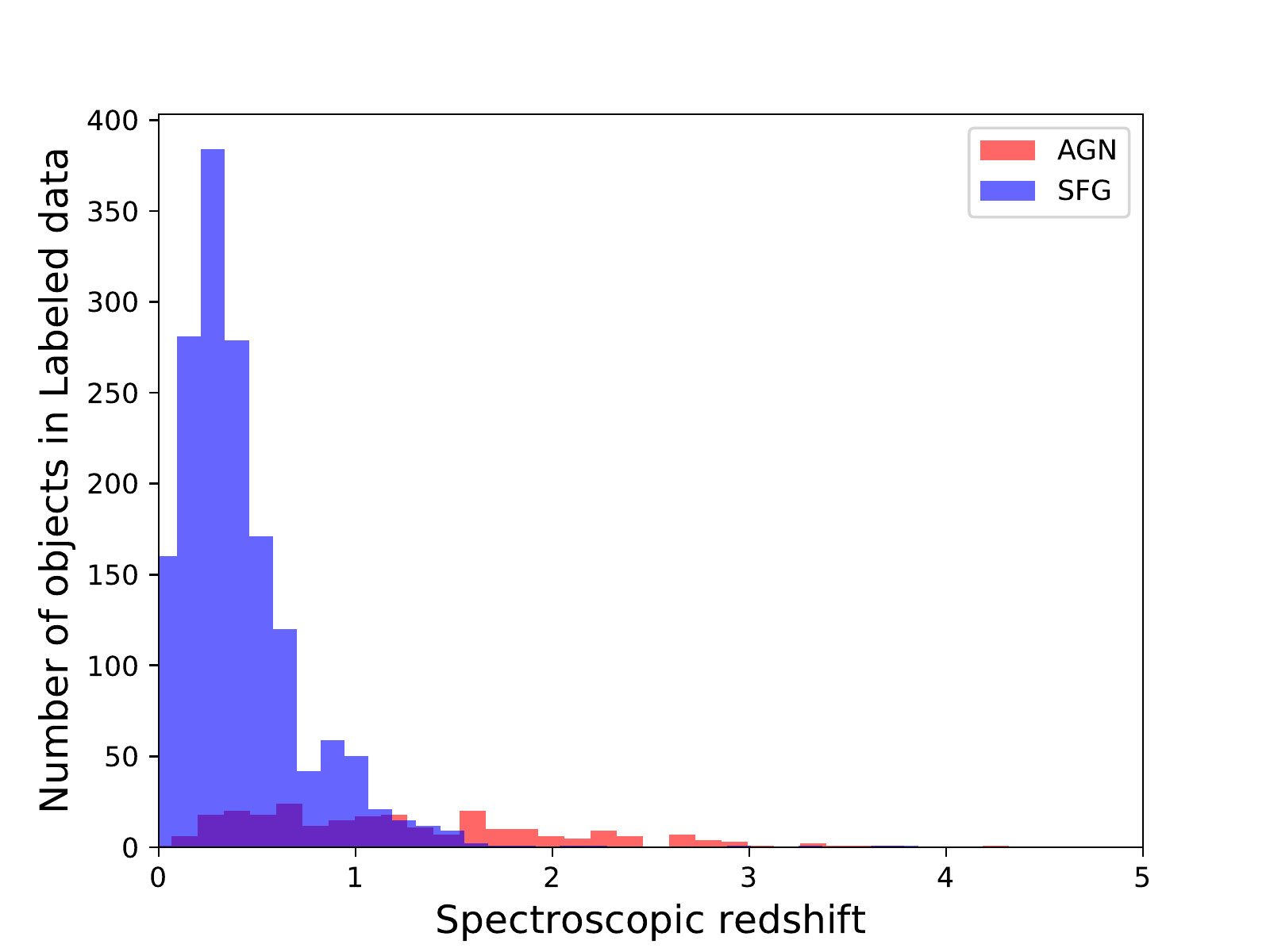}
    \caption{The spectroscopic redshift distribution of the labelled data.}
    \label{fig:figure4}
\end{figure}

\begin{figure*}
	\includegraphics[width=\textwidth]{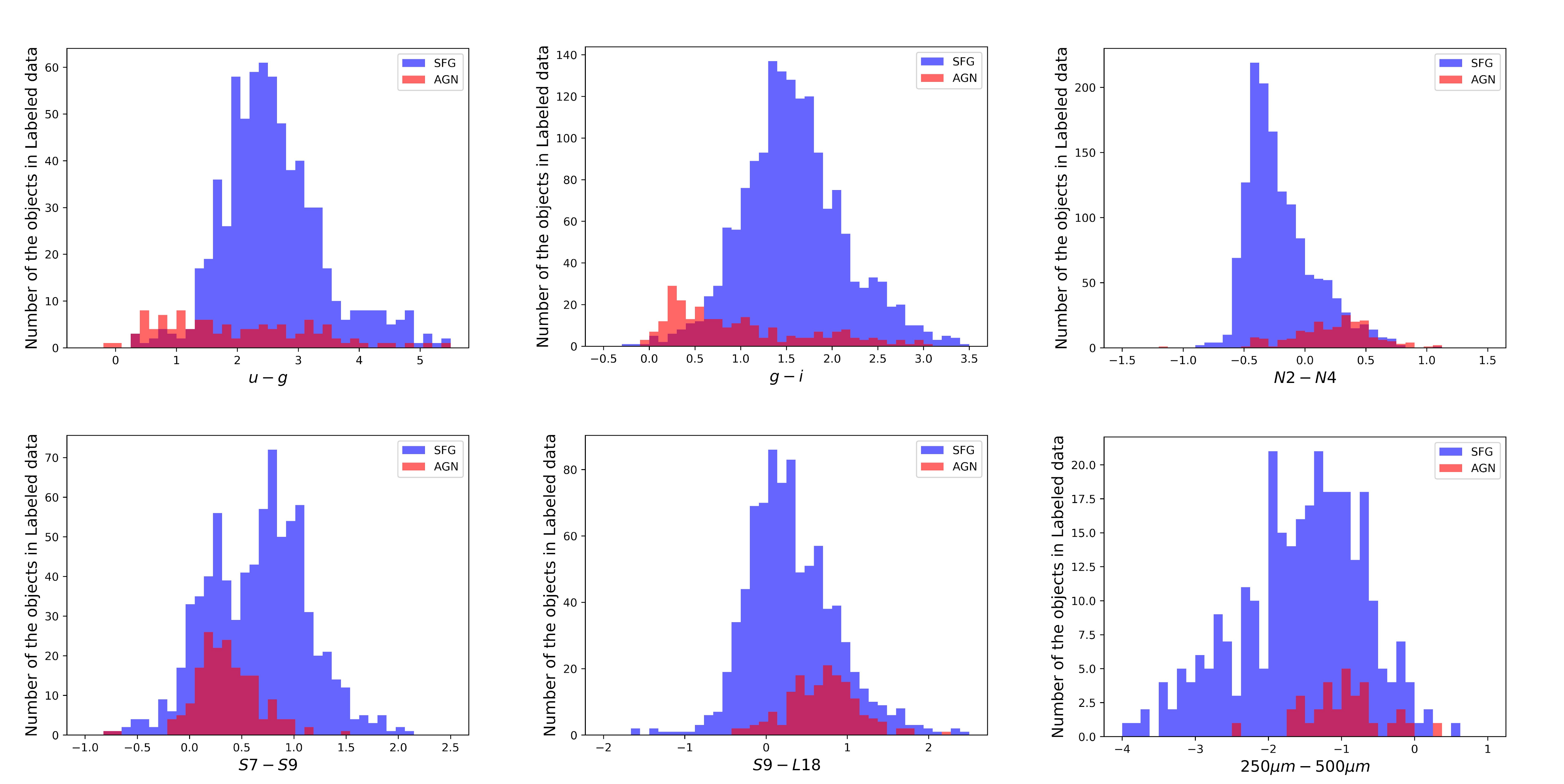}
    \caption{The colour distribution of the labelled data.  The upper left panel shows the distribution of $\emph{u-g}$, which represents the UV colour. The upper middle panel shows the optical colour distribution using $\emph{g-i}$.  The upper right panel is the NIR colour evaluated by $\emph{N2-N4}$.  The lower left and lower middle panel are both the colour distribution of MIR, composed by $\emph{S7-S9}$ and $\emph{S9-L18}$.  The lower right one is a FIR colour plot represented by $250\mu m-500\mu m$.}
    \label{fig:figure5}
\end{figure*}

The labelled data is comprised of 1870 objects, and these objects provide the foothold for NN training and validating.  In order to further understand the basic composition of our research, we plot the distribution of redshift and colour distribution based on the labelled objects.  The redshift distribution is obtained using the spectroscopic data mentioned in Sec.~\ref{sec:Sample selection}, shown in Fig.~\ref{fig:figure4}.  The colour distribution is evaluated using the Band-Merged catalogue mentioned in Sec.~\ref{sec:The 44 Band-Merged data}.  We separately give the plot of $\emph{u-g}$, $\emph{g-i}$, $\emph{N2-N4}$, $\emph{S7-S9}$, $\emph{S9-L18}$ and $250\mu m-500\mu m$, covering the distribution of UV, optical, NIR, MIR and FIR.  The 6 colour distributions are shown in Fig.~\ref{fig:figure5}.

\subsection{Data preprocessing}
\label{sec:Data preprocessing} 

In order to measure the performance of NN model correctly, we validate the model performance by K-fold cross validation (\citealt{Bishop2006}) with K = 10.  The labelled data are equally divided into 10 groups; whenever the training is performed, one of the groups (that is, 10\% of data) is excluded from the training to serve as validation data.  Sequential exclusion  and training would be repeated until all folds have once been the validation data.  We then take the performance average from all 10 trainings as our K-fold cross validation result.

All sources have at most 44 available band observations, and each observation has a pair of magnitude and magnitude error.  For each unavailable band observation, we fill in 0 instead.  Moreover, we conform that filling the missing data with the median value of the band or the median value of the neighbouring filters also give out the similar results.  We then reshape the magnitudes and magnitude errors to a $(44 \times 2)$ array (Fig.~\ref{fig:figure3}).  To make it more convenient to refer to other machine learning background papers, we trivially denote the array shape as $(44 \times 2 \times 1)$  in the following sections.

\subsection{Model Architecture}
\label{sec:Model Architecture}

\begin{figure*}
	\includegraphics[width=0.9\textwidth]{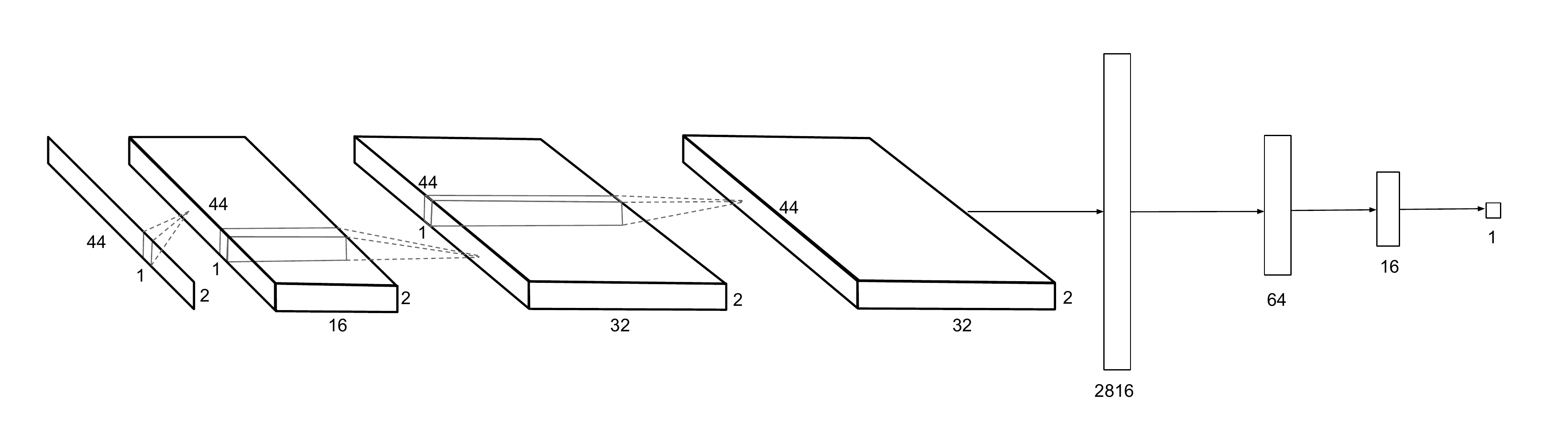}
    \caption{An illustration of the architecture of our NN. The network’s input is $(44 \times 2 \times 1)$, and the output feature map from each layer has a shape $(44 \times 2 \times 16)$, $(44 \times 2 \times 32)$, $(44 \times 2 \times 32)$.  It is then flattened to a vector with 2816 entries, and processed two fully-connected layers with 64 and 16 neurons, and finally output a single scalar.}
    \label{fig:figure6}
\end{figure*}

We summarised the architecture of our NN model in Fig.~\ref{fig:figure6}.  The NN has 5 learned layers, including 3
convolutional layers and 2 fully-connected layers. In the following subsections, we describe the overall architecture of the model and the specific technique we used during training.  

\subsubsection{Convolutional layer}
\label{sec:Convolutional layer} 

As described in Section~\ref{sec:Data preprocessing}, our input feature map is a $(44 \times 2 \times 1)$ array.  Three convolutional layers are placed sequentially to capture the features of the photometry, and between each layer a Rectified Linear Unit (ReLU) function is used to perform a non-linear transform (\citealt{Krizhevsky2017}).  All the kernels of the convolutional layers have the size $(1 \times 2)$.  In each layer, respectively, 16, 32, 32 kernels are used to capture the band features. In addition, padding was used to maintain the size of feature map.  Thus, the output feature map from each layer has a shape $(44 \times 2 \times 16)$, $(44 \times 2 \times 32)$, $(44 \times 2 \times 32)$, respectively.  In addition, we apply batch normalisation (\citealt{ioffe2015batch}), a method to re-centre and re-scale data in each layer, to enhance the training and apply L2 regularisation (\citealt{Cortes2012}), a method which adds NN weighting information in the loss function,  to avoid overfitting. 

\subsubsection{Fully-connected layer}
\label{sec:Fully-connected layer} 

The final output feature map of the convolutional layers is a $(44 \times 2 \times 32)$ array.  It is then flattened to a vector with 2816 entries and entered into fully-connected layers.  Two fully-connected layers are placed, featuring with 64 and 16 neurons respectively. ReLU function is also used between each layer.  In addition, we apply L2 regularisation and dropout (\citealt{Srivastava2014}), a method which disable a portion of units in NN during training, to avoid overfitting.

The output of the last layer is immediately summed and mapped by a sigmoid function.  This operation ensure that the NN outputs a single scalar range from 0 to 1.

\subsubsection{Focal Loss}
\label{sec:Focal Loss} 

Usually, cross-entropy is applied as the loss function of NN. The algorithm optimise the trainable parameter based on the first order differential of such function.  We denote $y$ as the ground truth of the sample, where 0 represents SFG and 1 represents AGN, and denote $p$ as the single scalar output of NN, where it ranges from 0 to 1, then the cross-entropy loss function is written as:

\begin{equation}
\ \ \ \ \ \ Loss_{\scriptscriptstyle CE}=-y \log{p} - (1 - y)\log{(1 - p)}
\label{eq:crossentropy}
\end{equation}

Note that when the ground truth and the NN output are highly consistent (ex. $(y, p) = (0, 0.01)$, or $(y, p) = (1, 0.99)$), Eq.~(\ref{eq:crossentropy}) is very close to -1.  On the other hand, if the results are not consistent (ex. $(y, p) = (0, 0.95)$, or $(y, p) = (1, 0.07)$), Eq.~(\ref{eq:crossentropy}) is more close to 0, making the loss larger.  The purpose of NN training is to decrease the value of this loss function so that the NN output is consistent with the ground truth.

However, Eq.~(\ref{eq:crossentropy}) performs poorly in our case.  The reason is that in our training sample there are only roughly 10\% AGNs. Such a fact causes an unavoidable bias on AGN recognition——the large population of SFGs in the training set leads the NN more likely to classify an AGN to be a SFG, while our main purpose is to identify AGNs from SFGs.  If we naïvely apply cross-entropy on training, the AGN completeness eventually fall under 50 \%.

In order to avoid such problem, we instead use focal loss (\citealt{Lin2017}), which is a modified cross-entropy loss function for up-weighting the hard training samples, written as

\begin{equation}
\label{eq:FC loss}
Loss_{\scriptscriptstyle FL}=-y\alpha {(1 - p)}^{\gamma} \log{p} - (1 - y)(1 - \alpha) {p}^{\gamma}\log{(1 - p)}, 
\end{equation}

where $\alpha\in[0, 1]$ and $\gamma > 0$.
By choosing a larger $\alpha$, the weighting of missing an AGN is enlarged.  Moreover, the base of $\gamma$ ($1 - p$ in the first term and $p$ in the second term) is the difference between the NN scalar output and the true answer; thus, choosing a larger $\gamma$ gives those worse-performing cases an exponentially larger weighting.

\subsubsection{Training the Neural Network}
\label{sec:Training the Neural Network} 

The procedure of NN training is illustrated in Algorithm.~\ref{alg:Training}, and we make use of the deep learning framework Keras\footnote{https://keras.io/} to implement it.  We use Adam optimisation (\citealt{Kingma2015}), an adaptive learning rate optimisation algorithm, to improve the NN by cycling "Input training data -- Evaluate loss -- Adam optimisation". We denote cycling it one time as an "epoch".

The training epochs come to the end when a specific condition is satisfied. This termination condition generally could be written as 

\begin{equation}
\label{eq:monitor}
\left\{
             \begin{array}{lr}
             (monitor)_{t} - (monitor)_{t-M} < \delta \\
             monitor = N * (AGN\ completeness) + accuracy,  \\
             \end{array}
\right.
\end{equation}

where 
$AGN\ completeness$ and $accuracy$ are defined in Section~\ref{sec:The Neural-Network performance},
$t$ denote the epoch,
$N$ indicates the weighting in the termination condition, 
$M$ indicates the epoch of waiting the monitor not improving, and   
$\delta$ indicates the minimum change of the $monitor$ that could be qualified as improvement.

Intuitively, Eq.~(\ref{eq:monitor}) is a trick that trace the improvement of NN performance and terminate the training automatically before overfitting occurred. 

It should be emphasised that the AGN completeness and accuracy is based on validation set.  The validation set data do not participate in training but it helps us decide when to stop training.  

The epoch of training can not be predetermined because the AGN completeness drops drastically if the training last too long.  Thus, we need to carefully set up this termination condition.  This trick assures the NN's performance while facing the real world condition.

\begin{algorithm*}
  \caption{Training algorithm for AGN recognition model}
  \label{alg:Training}
  \begin{algorithmic}[1]
  \State Input the training set and the validation set data. \Comment{Fig.~\ref{fig:figure2}, left down; Fig.~\ref{fig:figure3};  Section.~\ref{sec:Sample selection}}
  \State Initialise the weights of deep neural network. \Comment{Fig.~\ref{fig:figure6}; Section.~\ref{sec:Model Architecture}}
  \For{$iteration=1,2\ldots$}
    \State Perform AGN/SFG classification on the training set by deep neural network.
    \State Evaluate the Focal Loss of the last stage. \Comment{Eq.~\ref{eq:FC loss}}
    \State Perform Adam optimisation on the deep neural network.
    \State Evaluate the $monitor$ value on the validation set and record. \Comment{Eq.~\ref{eq:monitor}, lower}
    \If{termination condition satisfied} \Comment{Eq.~\ref{eq:monitor}, upper}
        \State \textbf{break} \Comment{Training Complete}
    \EndIf
  \EndFor
  \State Save the weights of deep neural network.
  \end{algorithmic}
\end{algorithm*}

\section{Empirical Result}
\label{sec:Empirical Result}
\begin{figure*}
	\includegraphics[width=\textwidth]{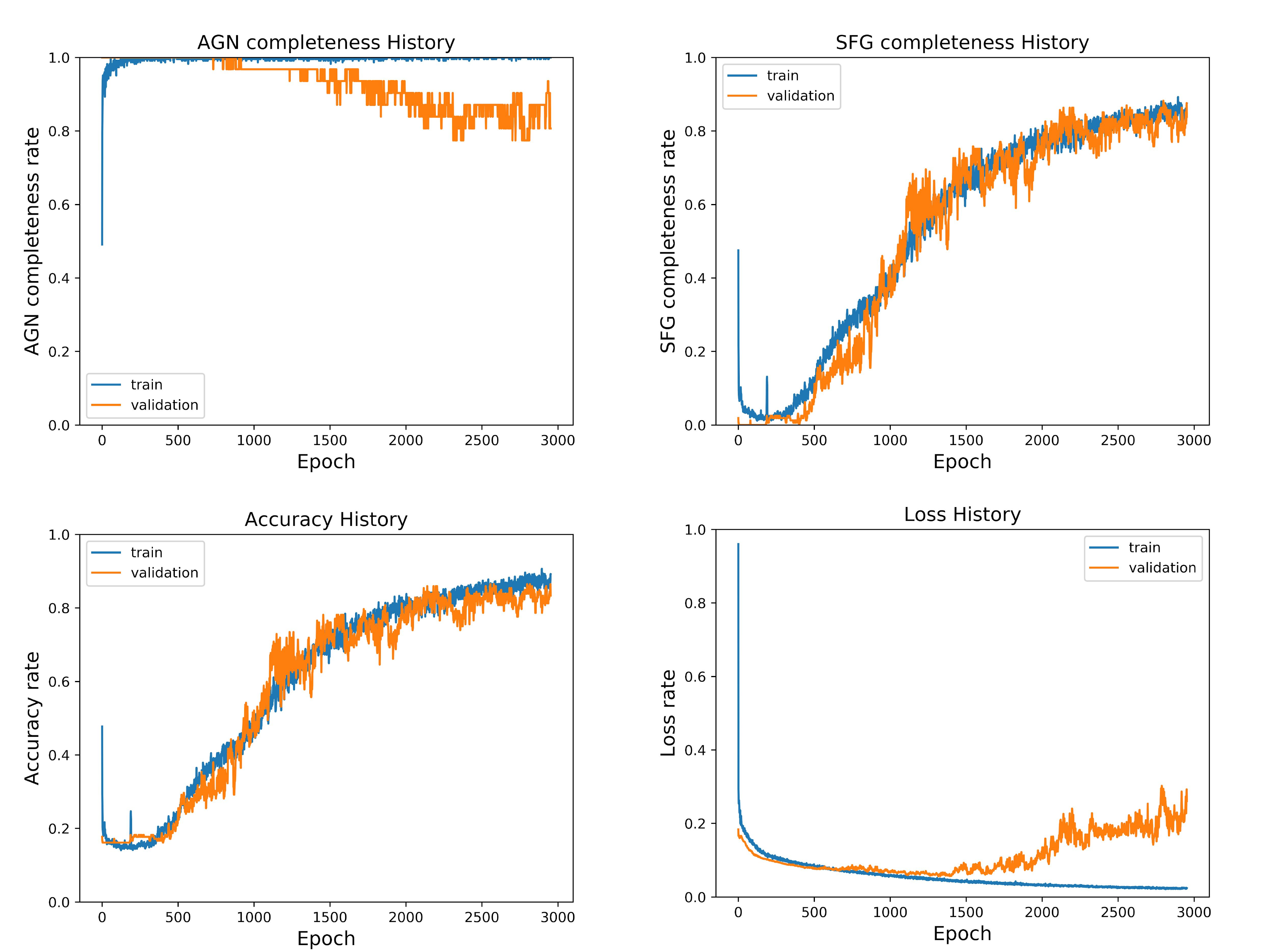}
    \caption{An example of NN training history, including the evolution of the AGN completeness (upper left panel), SFG completeness (upper right panel), accuracy (lower left panel) and focal loss (lower right panel). The AGN completeness increases from 30\% to almost 100\% in only 300 training epochs, and slightly decreases after 1500 epochs.  The accuracy and SFG completeness are gradually raised until~3000 epochs.  The focal loss descends in a stable manner before 1500 epochs and increases again because it is sensitive to the decline of AGN completeness.}
    \label{fig:figure7}
\end{figure*}

The result of the training is highly stochastic.  A small difference of hyperparameters (the parameters configuring the NN models) could lead to a totally different outcome.  Even if the hyperparameters are the same, the outcome will also not be the same between two tries; this is because the trainable parameters of NN are initialised randomly.  In our experiments, most of the time, we need to repeat the training several times to obtain the best performance of that set of hyperparameters. In the following sections, we provide some of the well-performing results and their corresponding hyperparameters.  The hyperparameters are chosen by hand tuning, which means we start from an arbitrary set of hyperparameters and manually, sequentially adjust it.

\subsection{The Neural-Network training history}
\label{sec:The Neural-Network training history}
Fig.~\ref{fig:figure7} shows an example of NN training history, including the evolution of the AGN completeness, accuracy and Focal loss.  We can see the AGN completeness increased from 50\% to almost 100\% in only 200 training epochs, however, at the cost of a very low accuracy and SFG completeness.  This result is due to the large setting of $\alpha$ in (2), which is $\alpha = 0.99$.  This setting induces an effect  like each AGN in the training set has a weighting 100 times larger than each SFG; thus the NN is more likely to classify an object as an AGN.  Fortunately, as the training progressed, the accuracy and SFG completeness is gradually increased.  The AGN completeness might slightly drop at this stage and the focal loss would sensitively response to the decline of AGN completeness since the focal loss is dominated by the large setting of $\alpha$. Usually, this status is held before the training comes to 1500\textasciitilde3000 epochs.  At the end of this stage, we reach the optimal point of this training.  If the hyperparameters are set properly, at this optimal point the accuracy would be at least above 80\% with the AGN completeness staying above 85\%.  If we keep training the NN, making the training steps far away from the optimal point, the accuracy will still be raised, but AGN completeness in validation set would drop drastically, making the training meaningless.  Thus, the termination condition mentioned in Section~\ref{sec:Training the Neural Network} helps us automatically stop the training near the optimal point.  

\begin{table}
	\centering
	\caption{The hyperparameters (defined in Eq.~\ref{eq:FC loss}, Eq.~\ref{eq:monitor}) and the performance of the NN models under K-fold cross validation.  }
	\label{tab:table1}
	\begin{tabular}{lll} 
		\hline
		\  & Hyperparameters & K-fold cross validation performance\\
		\hline
		Model A & $\alpha = 0.99$ &  AGN completeness $= 85.42\%$ \\
		\       & $\gamma = 2$    &  SFG completeness $= 85.09\%$ \\
		\       & $M = 4$         &  Accuracy $= 85.14\%$ \\
		\       & $N = 1.5$       &  ROC AUC $= 88.38\%$ \\
		\       & $\delta = -0.10$& \ \\
		\hline
		Model B & $\alpha = 0.99$ &  AGN completeness $= 85.83\%$ \\
		\       & $\gamma = 2$    &  SFG completeness $= 83.12\%$ \\
		\       & $M = 4$         &  Accuracy $= 83.60\%$ \\
		\       & $N = 3$         &  ROC AUC $= 88.89\%$  \\
		\       & $\delta = -0.15$& \ \\
		\hline
		Model C & $\alpha = 0.99$ &  AGN completeness $= 86.43\%$ \\
		\       & $\gamma = 2$    &  SFG completeness $= 81.17\%$ \\
		\       & $M = 6$         &  Accuracy $= 82.10\%$ \\
		\       & $N = 5$         &  ROC AUC $= 88.56 \%$  \\
		\       & $\delta = -0.20$& \ \\
		\hline
	\end{tabular}
\end{table}

\begin{figure}
	\includegraphics[width=\columnwidth]{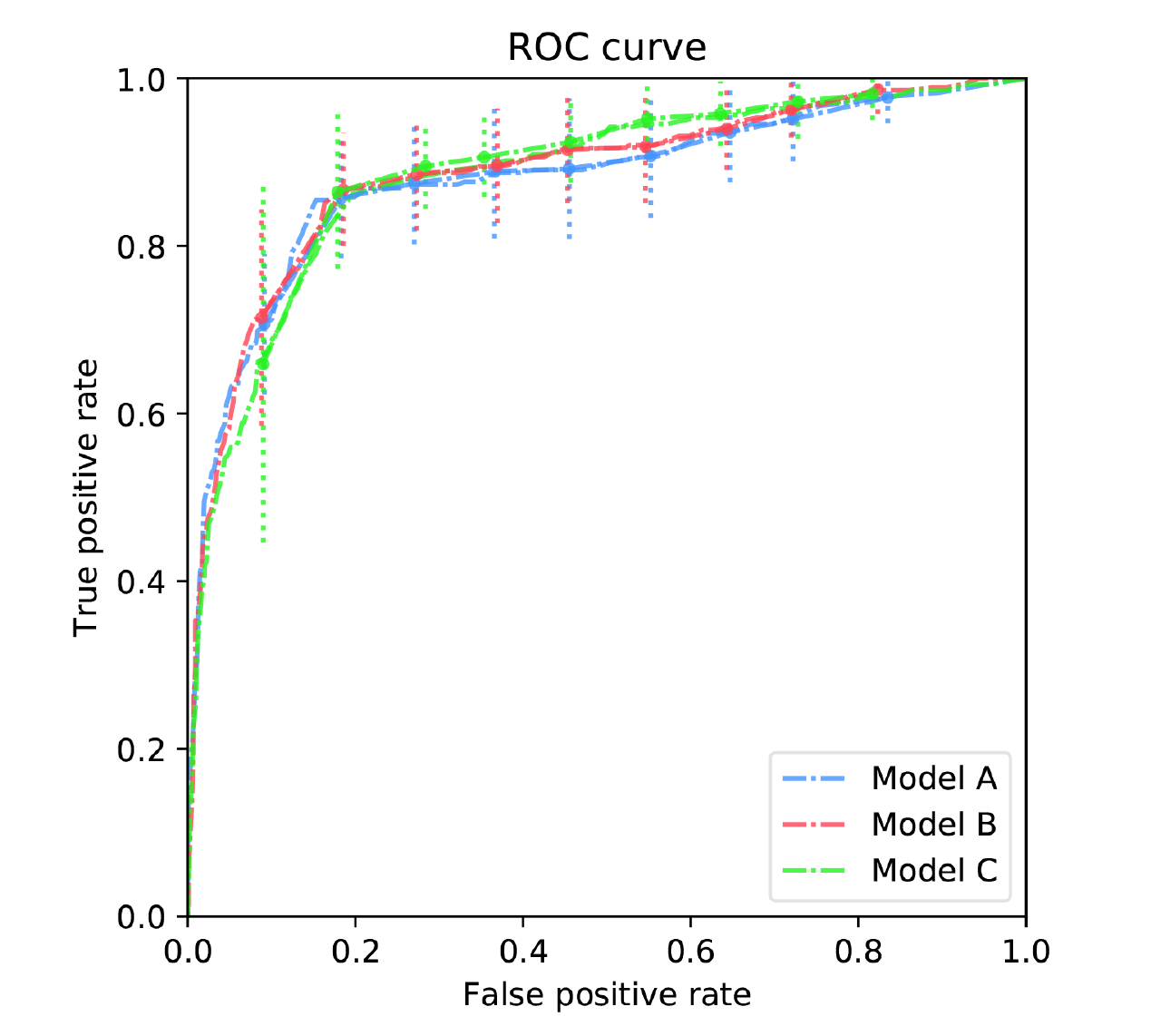}
    \caption{The ROC curves of the NN models referred in Table.~\ref{tab:table1}.  The result here is also using K-fold cross validation.}
    \label{fig:figure8}
\end{figure}

\begin{figure}
	\includegraphics[width=\columnwidth]{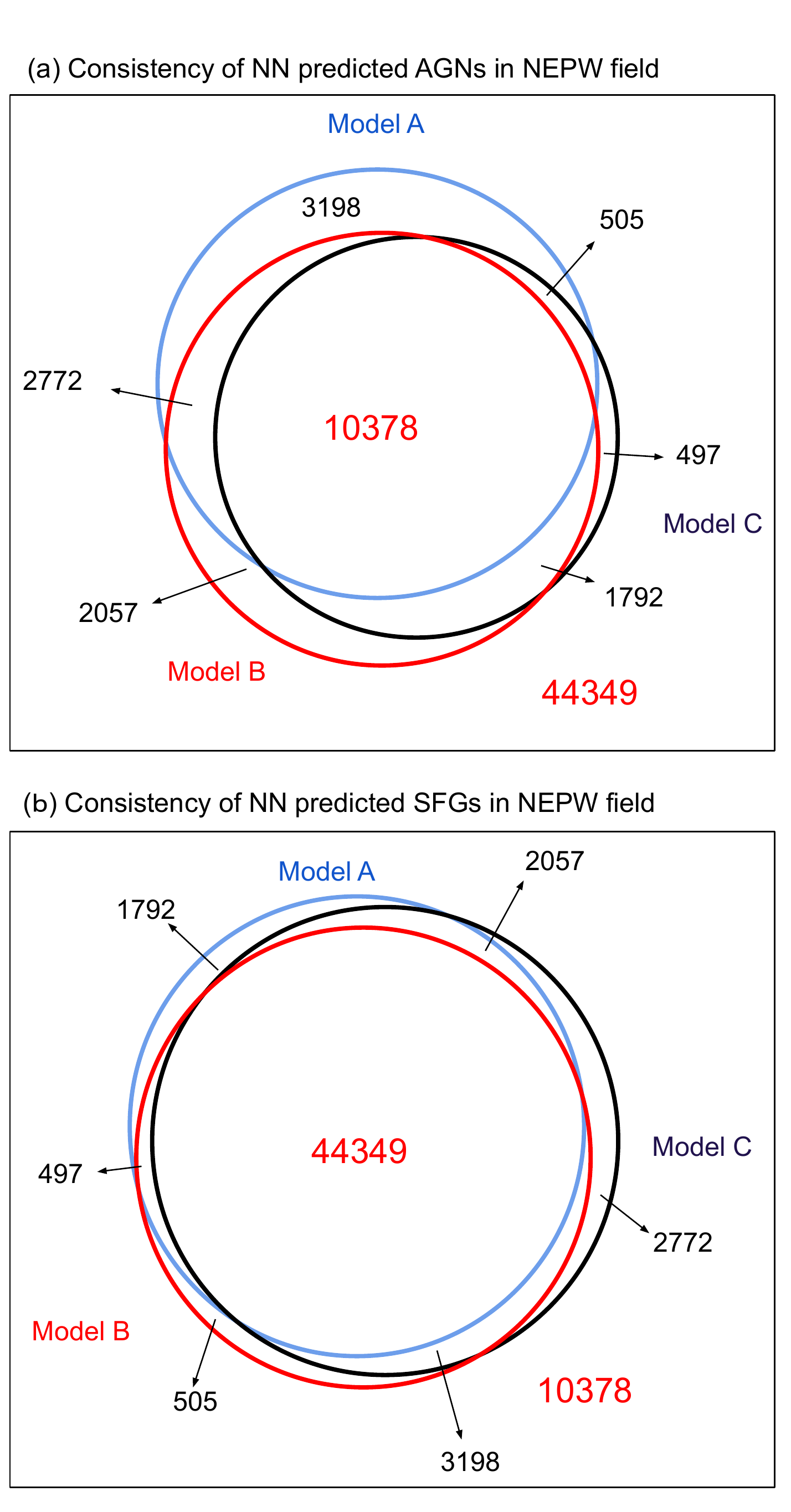}
    \caption{The consistency of the inference result from the three models in Table.~\ref{tab:table1}.  The area implies the number of objects, but not drawn to scale.  Upper panel shows the overlap of the AGN prediction; lower panel shows the overlap of the SFG prediction.  The SFG prediction has a higher consistency.}
    \label{fig:figure9}
\end{figure}

\begin{figure}
	\includegraphics[width=\columnwidth]{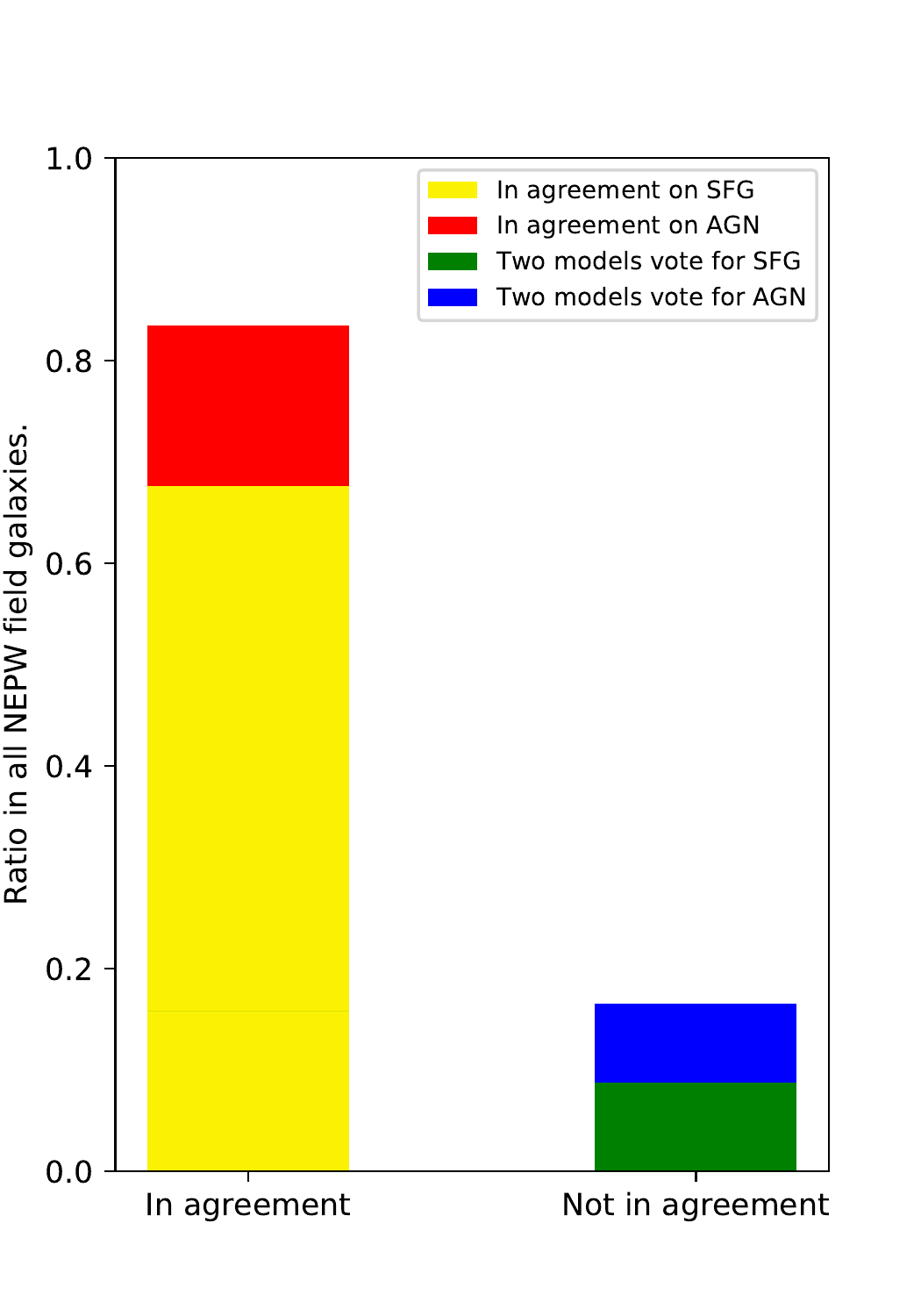}
    \caption{The bar graph of Fig.~\ref{fig:figure9}.  83.5\% NEPW field objects receive consistent result from the three models.  These consistent results are accumulated and present at the left side.}
    \label{fig:figure10}
\end{figure}

\subsection{The Neural-Network performance on AGN recognition}
\label{sec:The Neural-Network performance}
We eventually save the best-performing NN models after the training and record its validation set performance.  The performance is validated using K-fold cross validation (\citealt{Bishop2006}), with a total of 10 folds (K=10).

We use a total of four metrices to present the performance of AGN recognition models in our work.  These metrices are AGN completeness, SFG completeness, accuracy and area under the curve of receiver operating characteristic (ROC AUC).  

AGN completeness is defined as:
\begin{equation}
AGN\ \ completeness = True\ \ positive\ \ rate = \frac{TP}{TP + FN}, 
\end{equation}

where $TP (True\ positive)$ denote the number of AGNs correctly identified by the model, and $FN (False Negative)$ denote the number of AGNs incorrectly excluded by the model.

SFG completeness is defined as:
\begin{equation}
SFG\ \ completeness = True\ \ negative\ \ rate = \frac{TN}{TN + FP}, 
\end{equation}

where $TN (True\ Negative)$ denote the number of SFGs correctly identified by the model, and $FP (False\ Positive)$ denote the number of SFGs incorrectly excluded by the model.

Accuracy is defined as:
\begin{equation}
Accuracy = \frac{TP + TN}{Total}, 
\end{equation}

where $Total$ denote the number of all objects in validation set.

ROC AUC (\citealt{Bradley1997}) is defined as the area under the ROC curve, which is the plot created by plotting true positive rate (TPR) against false positive rate (FPR) at various threshold settings.  False positive rate is defined as:
\begin{equation}
False\ \ positive\ \ rate = \frac{FP}{TN + FP}, 
\end{equation}

The NN determines the AGN/SFG candidates by its single scalar output and a specific threshold.  With the various TPR and FPR results from various thresholds ranging from 0 to 1, we obtain the ROC curve of our NN model.  ROC curve provides a straight forward comparison between different classifiers.  The larger the AUC is, the better the model is.

Several hyperparameter sets and the K-fold cross validation results of these settings are shown in Table.~\ref{tab:table1}, and the corresponding ROC curves are shown in Fig.~\ref{fig:figure8}.  In Table.~\ref{tab:table1} we see that there is a trade off between the accuracy, SFG completeness and AGN completeness.   If the NN achieve an AGN completeness up to 86.43\%, then the accuracy and SFG completeness is about 82.10\% and 81.17\%, respectively; if the NN only cover about 85.42\% AGNs, then the accuracy reach 85.14\% and the SFG completeness comes to 85.09\%.  The result shows that NN model typically carry out an AGN recognition performance around 85\% level.  Furthermore, we show the comparison with traditional statistical analysis in Section~\ref{sec:Compare with SED fitting}, stating that NN could provide a more reliable way on AGN recognition problem.

\section{Discussions}
\label{sec:discussion}

\subsection{The inference result on whole NEP field}
\label{sec:The inference result on whole NEP field}
After the NN is well-trained, we use it to classify arbitrary objects in the NEPW field.  As shown in Fig.~\ref{fig:figure2}, there are 65548 objects in the NEPW field with spectroscopic and/or photometric redshift measurement but no classification result yet available. We apply all three models referred in Table.~\ref{tab:table1} (only one fold in totally 10 folds here); the inference result of these NN models is shown in Table.~\ref{tab:table2}.  

Comparing these three models, we obtain the estimates of AGN fraction (ratio of number of AGNs to total number of galaxies) between ~25\% to 34\%.  Note that these evaluations of AGN fraction include the objects in the training data.  The AGNs recognised by model with the smallest AGN fraction (24.89\%) are almost covered by the remaining two models too (only 497 exceptions).  The remaining two models give out a quite similar AGN fraction (34.01\% - 34.40\%), but the identified AGNs have relatively larger differences (3198 and 2057 AGNs were only identified by models A and B, respectively).

We also compare the inference result among these three models.  We show the result in Fig.~\ref{fig:figure9} and Fig.~\ref{fig:figure10}. The result shows that 83.49\% objects are receiving the same results from all three NN models, 8.78\% objects are voted as AGNs by one NN model and as SFGs by two NN models, 7.73\% objects are voted as AGNs by two NN models and as SFGs by one NN model.  Three NN models are showing high consistency in recognising SFGs, while the consistency is relatively low in recognising AGNs.  This difference might have resulted from the fact that the population of SFGs in the training data is larger than AGNs, thus the SFG information provided to the NN model is comparatively sufficient.
\begin{table}
	\centering
	\caption{The NEP field inference result of the three NN models in Table.~\ref{tab:table1}.  The estimations of AGN fraction include the population of training data.}
	\label{tab:table2}
	\begin{tabular}{rcccc} 
		\hline
		\ & Model A & Model B & Model C & training data\\
		\hline
		AGN:          & 16853   & 16999   & 13172   & 255   \\
		\specialrule{0em}{1pt}{1pt}
		SFG:          & 48695   & 48549   & 52376   & 1615   \\
		\specialrule{0em}{1pt}{1pt}
		Total:        & 65548   & 65548   & 65548   & 1870   \\
		\specialrule{0em}{3pt}{3pt}
		AGN fraction :& 34.01\% & 34.40\% & 24.89\% & \      \\
		\hline
	\end{tabular}
\end{table}

\subsection{Comparison with SED fitting result}
\label{sec:Compare with SED fitting}

\begin{figure}
	\includegraphics[width=1.0\columnwidth]{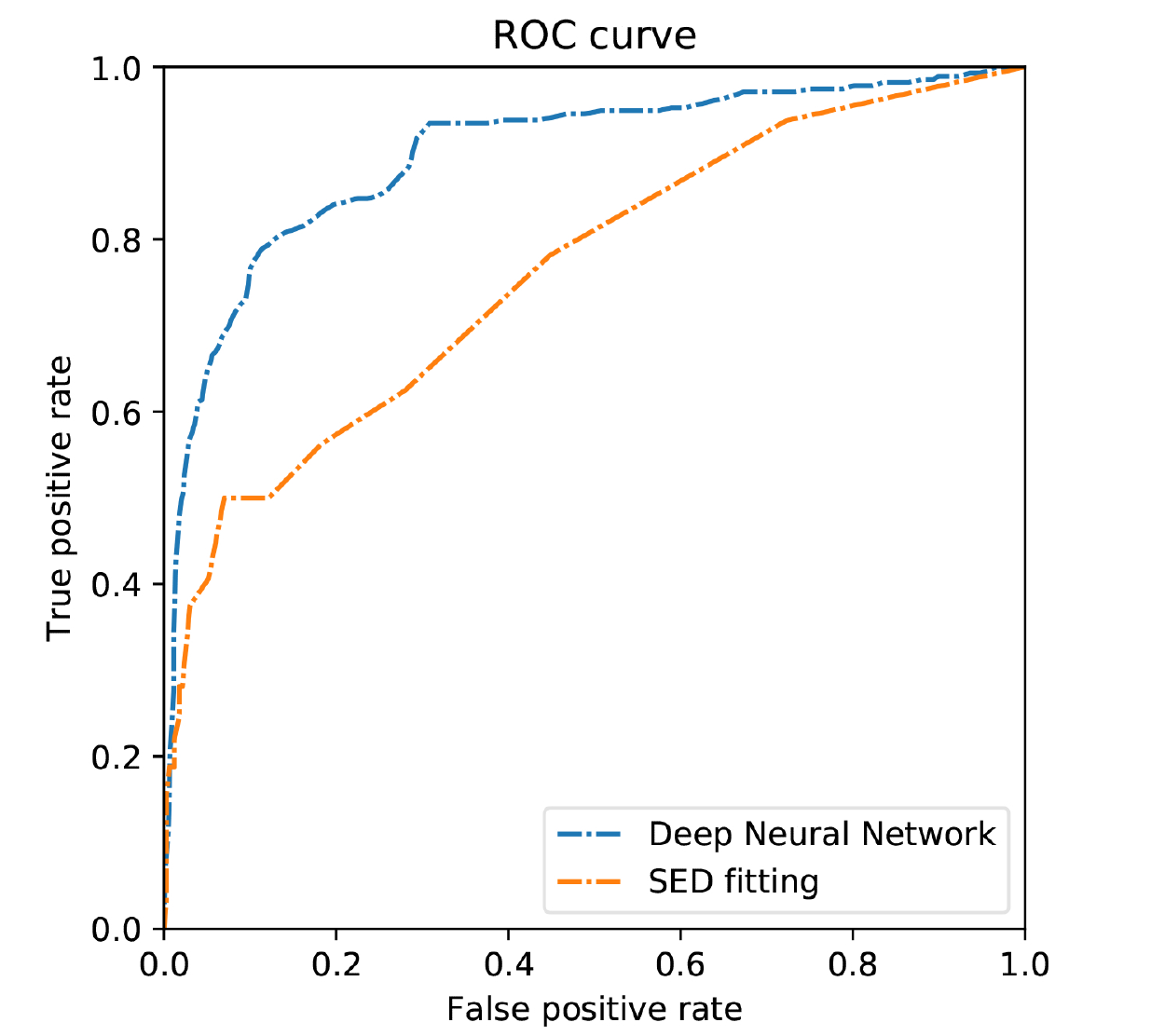}
    \caption{The comparison between NN model and SED fitting via ROC curve. The SED fitting result is using \begin{footnotesize}CIGALE\end{footnotesize}, provided by \citet{Wang2020}.  We show that the ROC AUC of NN model ($89.91\%$) is larger than SED fitting one ($76.23\%$), indicating that NN model is a better classifier.}
    \label{fig:figure11}
\end{figure}

We compare the NN performance with the SED fitting performance (\citealt{Wang2020}) by presenting the ROC curve of two methods.  In this comparison, NN metrices are using K-fold cross validation results from the validation sets, and the model is using the Model B referred in Table.~\ref{tab:table1}, and SED fitting metrices are validated by  the intersection of the labelled data shown in Fig.~\ref{fig:figure2} and the SED fitting applicable candidates.  The fitting model is provided by \begin{footnotesize}CIGALE\end{footnotesize}.  In this SED fitting work, the IR luminosity contribution of the galactic nucleus ($ f_{\rm AGN\_IR}$) is derived; a galaxy is identified as an AGN when $ f_{\rm AGN\_IR} \geq 0.2$.  Thus, a ROC curve of SED fitting model could be carried out by varying the $ f_{\rm AGN\_IR}$ threshold. We plot the ROC curve of our result and the result from \citet{Wang2020} in Fig.~\ref{fig:figure11}.  The ROC AUC of NN model and SED fitting are $89.91\%$ and $76.23\%$,  respectively, indicating that the NN model provides a more accurate selection compared with SED fitting.  Furthermore, SED fitting method require some critical IR detection (e.g. AKARI 18W, Herschel Spire PSW or PACS in our case), and a well-fitted result.  These constraints have limited the applicable candidates to only 1671 objects in all NEP field; in contrast, NN models provide 65548 object classifications in total, covering almost the whole NEPW field.  Thus, based on our testing result in the NEPW field samples, we state that NN is a better solution when it comes to AGN recognition.  

However, \citet{Wang2020} provided a more sophisticated investigation on physical properties of AGNs. They derive the physical properties (e.g. AGN contribution, star formation rate, etc) from \begin{footnotesize}CIGALE\end{footnotesize}, which we cannot obtain from the NN model unless it is trained to provide it.  Thus, additional properties could be obtained if we select AGN using our NN model and perform physical property analysis using SED fitting technique.

Another photometric, machine learning-based AGN recognition is performed in \citealt{Poliszczuk2019}.  The algorithm they used is a fuzzy support vector machine (FSVM).  However, instead of selecting sources in the NEPW field, they focus on the NEPD field. Compare with our NEPW field source, NEPD field is with narrower area and fainter detection, thus making a comparison between their work and ours is not straightforward.  In spite of resulting from different data, an informal comparison (not shown in this paper) shows that FSVM has a similar ROC curve performance compared with our NN model.

\subsection{The contribution of different range of observations.}
\label{sec:The contribution of different range of observations.}
\begin{figure}
	\includegraphics[width=1.0\columnwidth]{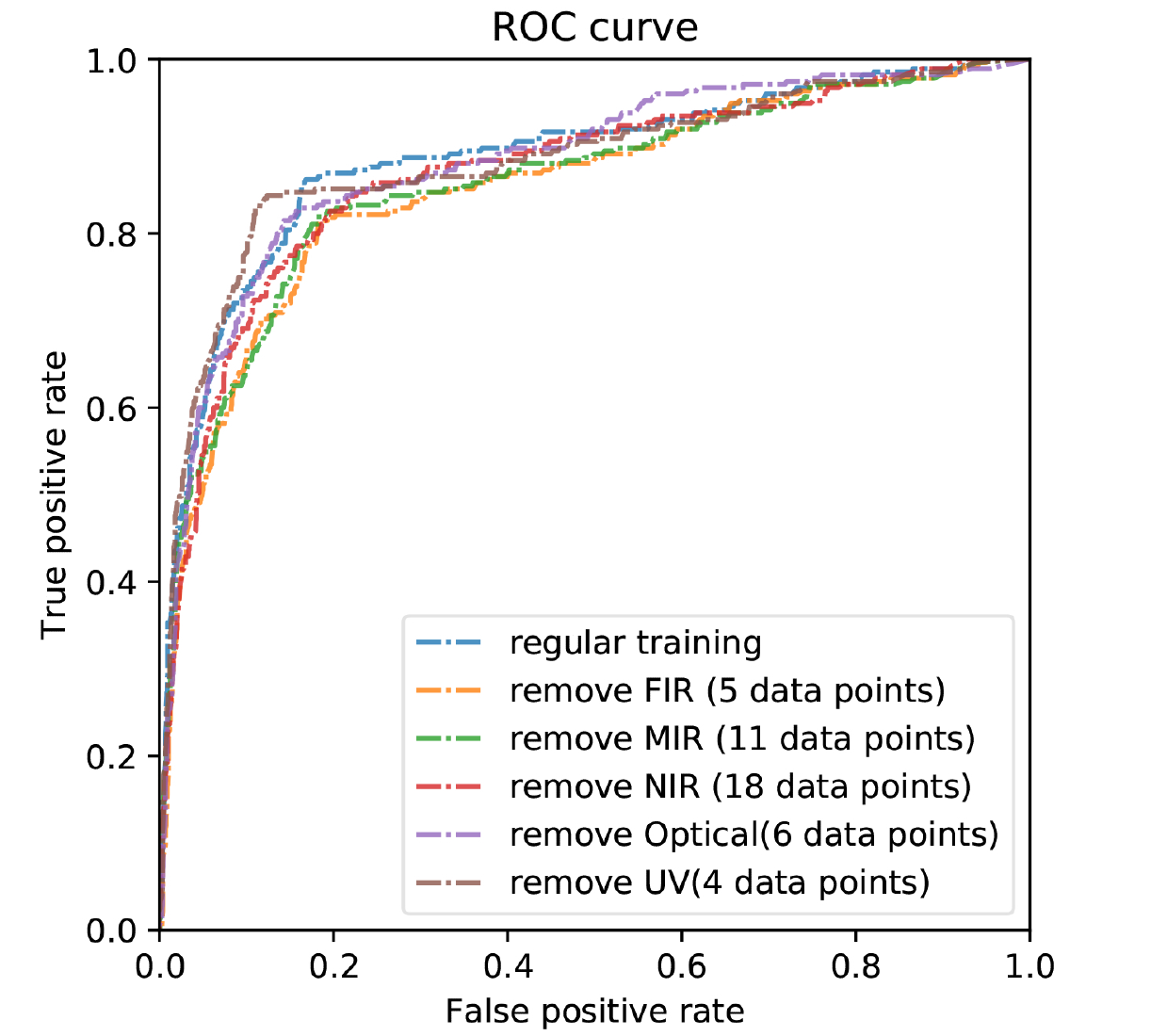}
    \caption{The ROC curve of the NN models, with some portion of bands are removed during training.  We remove 6 data points in FIR, 11 in MIR, 18 in NIR, 6 in optical and 4 in UV band for each training.  The result shows that removing FIR an MIR observations lead to slightly worse result, but the performances are not decreasing drastically.}
    \label{fig:figure12}
\end{figure}

In order to study the contribution of different ranges of observations in our training, we perform the experiments that train the NN under a constraint that a range of data points are removed.  Totally 6 experiments are performed in this part.  Except one regular training, we experimented removing FIR data (100-500 ${\mu}m$, 6 data points), MIR data (5.8-24 ${\mu}m$, 11 data points), NIR data (0.8-4.6 ${\mu}m$, 18 data points), Optical data (0.4-0.65 ${\mu}m$, 6 data points), and UV data (0.15-0.36 ${\mu}m$, 4 data points).  We show the training results of each experiment in the form of ROC curve in Fig.~\ref{fig:figure12}.  This set of experiments shows that removing FIR an MIR observations leads to slightly worse results, but the performance is not drastically decreasing.  Thus, based on this result, we can infer that none of FIR, MIR, NIR, Optical or UV observations are uniquely providing the key information for AGN recognition.

\section{Conclusions}
\label{sec:Conclusions}
A critical issue in the field of astrophysics is that although identifying AGNs from the normal SFGs is essential, in the NEPW field most of the objects are merely photometrically surveyed.  Not many X-ray or spectroscopic classification is available in this aspect, hence the AGNs in the NEPW field have not been well-identified yet.  In order to address such issues, we try a novel solution based on NN.  Eventually, our work resulted in three main conclusions:

\begin{itemize}
\item We verify that Deep Neural-Network is applicable in recognising AGNs using photometric data in the NEPW field, and gives out a feasible technique set.  The recognition accuracy, AGN completeness and SFG completeness are recorded to be around $82.10\% - 85.14\%$, $85.42\% - 86.43\%$ and $81.17\% - 85.09\%$,  respectively.
\item We publicly release a high-quality AGN/SFG classification catalogue covering the whole NEPW field based on Deep Neural-Network.  In this catalogue, $83.49\%$ of the galaxies have the same results from the three different Deep Neural-Network models, which differ on the hyperparameters.
\item We show that Deep Neural-Network provides a more reliable and less prerequisite classification result compared with the popular SED fitting method according to our testing samples in NEPW filed. As shown in the ROC AUC values of Deep Neural-Network and SED fitting method, the scores are $88.38\% - 88.89\%$ and $76.23\%$, respectively.
\end{itemize}

In summary, we provide a high-quality AGN/SFG classification catalogue in the NEPW field for immediate scientific use.  In addition, with the upcoming telescope in near future (e.g. JWST, Euclid, eROSITA, SPICA….etc), more and more training samples and photometrical bands would become available. We could consequently expect a further enhanced NN AGN recognition.

\section*{Acknowledgements}

We are very grateful to the anonymous referee for many insightful comments. This research is based on observations with $AKARI$, a JAXA project with the participation of ESA.  This research was conducted under the agreement on scientific cooperation between the Polish Academy of Sciences and the Ministry of Science and Technology (MOST) of Taiwan through grant 109-2927-I-007-505. TG acknowledges the support by the MOST of Taiwan through grant 108-2628-M-007 -004 -MY3.  TH is supported by the Centre for Informatics and Computation in Astronomy (CICA) at National Tsing Hua University (NTHU) through a grant from the Ministry of Education (MOE) of Taiwan.  AP and AP are supported by the Polish National Science Centre grant UMO-2018/30/M/ST9/00757 and by Polish Ministry of Science and Higher Education grant DIR/WK/2018/12.  TM is supported by UNAM-DGAPA PASPA and PAPIIT IN111319 as well as CONACyT 252531.

This work used high-performance computing facilities operated by CICA at NTHU. This equipment was funded by the MOE of Taiwan, MOST of Taiwan, and NTHU.

\section*{Data Availability}
The data is available upon request.


\bibliographystyle{mnras}
\bibliography{AIAGNR} 

\begin{thebibliography}{}
\makeatletter
\relax
\def\mn@urlcharsother{\let\do\@makeother \do\$\do\&\do\#\do\^\do\_\do\%\do\~}
\def\mn@doi{\begingroup\mn@urlcharsother \@ifnextchar [ {\mn@doi@}
  {\mn@doi@[]}}
\def\mn@doi@[#1]#2{\def\@tempa{#1}\ifx\@tempa\@empty \href
  {http://dx.doi.org/#2} {doi:#2}\else \href {http://dx.doi.org/#2} {#1}\fi
  \endgroup}
\def\mn@eprint#1#2{\mn@eprint@#1:#2::\@nil}
\def\mn@eprint@arXiv#1{\href {http://arxiv.org/abs/#1} {{\tt arXiv:#1}}}
\def\mn@eprint@dblp#1{\href {http://dblp.uni-trier.de/rec/bibtex/#1.xml}
  {dblp:#1}}
\def\mn@eprint@#1:#2:#3:#4\@nil{\def\@tempa {#1}\def\@tempb {#2}\def\@tempc
  {#3}\ifx \@tempc \@empty \let \@tempc \@tempb \let \@tempb \@tempa \fi \ifx
  \@tempb \@empty \def\@tempb {arXiv}\fi \@ifundefined
  {mn@eprint@\@tempb}{\@tempb:\@tempc}{\expandafter \expandafter \csname
  mn@eprint@\@tempb\endcsname \expandafter{\@tempc}}}

\bibitem[\protect\citeauthoryear{Alexander, Brandt, Hornschemeier, Garmire,
  Schneider, Bauer  \& Griffiths}{Alexander et~al.}{2001}]{Alexander2001}
Alexander D.~M.,  Brandt W.~N.,  Hornschemeier A.~E.,  Garmire G.~P.,
  Schneider D.~P.,  Bauer F.~E.,   Griffiths R.~E.,  2001, \mn@doi [The
  Astronomical Journal] {10.1086/323540}, 122, 2156

\bibitem[\protect\citeauthoryear{Baldwin, Phillips  \& Terlevich}{Baldwin
  et~al.}{1981}]{Baldwin1981}
Baldwin J.~A.,  Phillips M.~M.,   Terlevich R.,  1981, \mn@doi [Publications of
  the Astronomical Society of the Pacific] {10.1086/130766}, 93, 5

\bibitem[\protect\citeauthoryear{Bishop}{Bishop}{2006}]{Bishop2006}
Bishop C.~M.,  2006, Pattern Recognition and Machine Learning.
Springer-Verlag New York Inc., \url
  {https://www.ebook.de/de/product/5324937/christopher_m_bishop_pattern_recognition_and_machine_learning.html}

\bibitem[\protect\citeauthoryear{Bohlin, Colina  \& Finley}{Bohlin
  et~al.}{1995}]{Bohlin1995}
Bohlin R.~C.,  Colina L.,   Finley D.~S.,  1995, \mn@doi [The Astronomical
  Journal] {10.1086/117606}, 110, 1316

\bibitem[\protect\citeauthoryear{Bradley}{Bradley}{1997}]{Bradley1997}
Bradley A.~P.,  1997, \mn@doi [Pattern Recognition]
  {10.1016/s0031-3203(96)00142-2}, 30, 1145

\bibitem[\protect\citeauthoryear{Chabrier, Baraffe, Allard  \&
  Hauschildt}{Chabrier et~al.}{2000}]{Chabrier2000}
Chabrier G.,  Baraffe I.,  Allard F.,   Hauschildt P.,  2000, \mn@doi [The
  Astrophysical Journal] {10.1086/309513}, 542, 464

\bibitem[\protect\citeauthoryear{Chiu, Ho, Wang  \& Lai}{Chiu
  et~al.}{2020}]{Chiu2020}
Chiu Y.-L.,  Ho C.-T.,  Wang D.-W.,   Lai S.-P.,  2020, arXiv preprint
  arXiv:2007.06235

\bibitem[\protect\citeauthoryear{Collister \& Lahav}{Collister \&
  Lahav}{2004}]{Collister2004}
Collister A.~A.,  Lahav O.,  2004, \mn@doi [Publications of the Astronomical
  Society of the Pacific] {10.1086/383254}, 116, 345

\bibitem[\protect\citeauthoryear{Cortes, Mohri  \& Rostamizadeh}{Cortes
  et~al.}{2012}]{Cortes2012}
Cortes C.,  Mohri M.,   Rostamizadeh A.,  2012, arXiv preprint arXiv:1205.2653

\bibitem[\protect\citeauthoryear{Cybenko}{Cybenko}{1989}]{Cybenko1989}
Cybenko G.,  1989, Mathematics of control, signals and systems, 2, 303

\bibitem[\protect\citeauthoryear{De~Wei \& Yang}{De~Wei \&
  Yang}{2019}]{DeWei2019}
De~Wei K.~C.,  Yang A.,  2019, \mn@doi [{EPJ} Web of Conferences]
  {10.1051/epjconf/201920609006}, 206, 09006

\bibitem[\protect\citeauthoryear{{Ho} et~al.,}{{Ho} et~al.}{2020}]{Ho2020}
{Ho} S. C.-C.,  et~al., 2020, \mnras, in press

\bibitem[\protect\citeauthoryear{Hornik}{Hornik}{1991}]{Hornik1991}
Hornik K.,  1991, Neural networks, 4, 251

\bibitem[\protect\citeauthoryear{Huang, Goto, Hashimoto, Oi  \&
  Matsuhara}{Huang et~al.}{2017}]{Huang2017}
Huang T.-C.,  Goto T.,  Hashimoto T.,  Oi N.,   Matsuhara H.,  2017, \mn@doi
  [Monthly Notices of the Royal Astronomical Society] {10.1093/mnras/stx1947},
  471, 4239

\bibitem[\protect\citeauthoryear{Huang et~al.,}{Huang et~al.}{2020}]{Huang2020}
Huang T.-C.,  et~al., 2020, \mn@doi [Monthly Notices of the Royal Astronomical
  Society] {10.1093/mnras/staa2459}, 498, 609

\bibitem[\protect\citeauthoryear{Hwang et~al.,}{Hwang et~al.}{2007}]{Hwang2007}
Hwang N.,  et~al., 2007, \mn@doi [The Astrophysical Journal Supplement Series]
  {10.1086/519216}, 172, 583

\bibitem[\protect\citeauthoryear{Ilbert et~al.,}{Ilbert
  et~al.}{2008}]{Ilbert2008}
Ilbert O.,  et~al., 2008, \mn@doi [The Astrophysical Journal]
  {10.1088/0004-637x/690/2/1236}, 690, 1236

\bibitem[\protect\citeauthoryear{Ioffe \& Szegedy}{Ioffe \&
  Szegedy}{2015}]{ioffe2015batch}
Ioffe S.,  Szegedy C.,  2015, arXiv preprint arXiv:1502.03167

\bibitem[\protect\citeauthoryear{Jarrett et~al.,}{Jarrett
  et~al.}{2011}]{Jarrett2011}
Jarrett T.~H.,  et~al., 2011, \mn@doi [The Astrophysical Journal]
  {10.1088/0004-637x/735/2/112}, 735, 112

\bibitem[\protect\citeauthoryear{Jeon, Im, Ibrahimov, Lee, Lee  \& Lee}{Jeon
  et~al.}{2010}]{Jeon2010}
Jeon Y.,  Im M.,  Ibrahimov M.,  Lee H.~M.,  Lee I.,   Lee M.~G.,  2010,
  \mn@doi [The Astrophysical Journal Supplement Series]
  {10.1088/0067-0049/190/1/166}, 190, 166

\bibitem[\protect\citeauthoryear{Jeon, Im, Kang, Lee  \& Matsuhara}{Jeon
  et~al.}{2014}]{Jeon2014}
Jeon Y.,  Im M.,  Kang E.,  Lee H.~M.,   Matsuhara H.,  2014, \mn@doi [The
  Astrophysical Journal Supplement Series] {10.1088/0067-0049/214/2/20}, 214,
  20

\bibitem[\protect\citeauthoryear{Juneau, Dickinson, Alexander  \& Salim}{Juneau
  et~al.}{2011}]{Juneau2011}
Juneau S.,  Dickinson M.,  Alexander D.~M.,   Salim S.,  2011, \mn@doi [The
  Astrophysical Journal] {10.1088/0004-637x/736/2/104}, 736, 104

\bibitem[\protect\citeauthoryear{Juneau et~al.,}{Juneau
  et~al.}{2013}]{Juneau2013}
Juneau S.,  et~al., 2013, \mn@doi [The Astrophysical Journal]
  {10.1088/0004-637x/764/2/176}, 764, 176

\bibitem[\protect\citeauthoryear{Kim et~al.,}{Kim et~al.}{2012}]{Kim2012}
Kim S.~J.,  et~al., 2012, \mn@doi [Astronomy {\&} Astrophysics]
  {10.1051/0004-6361/201219105}, 548, A29

\bibitem[\protect\citeauthoryear{Kim et~al.,}{Kim et~al.}{2020}]{Kim2020}
Kim S.~J.,  et~al., 2020, \mn@doi [Monthly Notices of the Royal Astronomical
  Society] {10.1093/mnras/staa3359}, 500, 4078

\bibitem[\protect\citeauthoryear{Kingma \& Ba}{Kingma \& Ba}{2019}]{Kingma2015}
Kingma D.~P.,  Ba J.~A.,  2019, arXiv preprint arXiv:1412.6980, 434

\bibitem[\protect\citeauthoryear{Krizhevsky, Sutskever  \& Hinton}{Krizhevsky
  et~al.}{2017}]{Krizhevsky2017}
Krizhevsky A.,  Sutskever I.,   Hinton G.~E.,  2017, \mn@doi [Communications of
  the {ACM}] {10.1145/3065386}, 60, 84

\bibitem[\protect\citeauthoryear{Krumpe et~al.,}{Krumpe
  et~al.}{2014}]{Krumpe2014}
Krumpe M.,  et~al., 2014, \mn@doi [Monthly Notices of the Royal Astronomical
  Society] {10.1093/mnras/stu2010}, 446, 911

\bibitem[\protect\citeauthoryear{Lacy et~al.,}{Lacy et~al.}{2004}]{Lacy2004}
Lacy M.,  et~al., 2004, \mn@doi [The Astrophysical Journal Supplement Series]
  {10.1086/422816}, 154, 166

\bibitem[\protect\citeauthoryear{Lee et~al.,}{Lee et~al.}{2009}]{Lee2009}
Lee H.~M.,  et~al., 2009, \mn@doi [Publications of the Astronomical Society of
  Japan] {10.1093/pasj/61.2.375}, 61, 375

\bibitem[\protect\citeauthoryear{Lin, Goyal, Girshick, He  \& Dollar}{Lin
  et~al.}{2017a}]{Lin2017}
Lin T.-Y.,  Goyal P.,  Girshick R.,  He K.,   Dollar P.,  2017a, in 2017 {IEEE}
  International Conference on Computer Vision ({ICCV}). {IEEE},
  \mn@doi{10.1109/iccv.2017.324}

\bibitem[\protect\citeauthoryear{Lin, Tegmark  \& Rolnick}{Lin
  et~al.}{2017b}]{Lin2017a}
Lin H.~W.,  Tegmark M.,   Rolnick D.,  2017b, \mn@doi [Journal of Statistical
  Physics] {10.1007/s10955-017-1836-5}, 168, 1223

\bibitem[\protect\citeauthoryear{Lu, Pu, Wang, Hu  \& Wang}{Lu
  et~al.}{2017}]{Lu2017}
Lu Z.,  Pu H.,  Wang F.,  Hu Z.,   Wang L.,  2017, Advances in neural
  information processing systems, 30, 6231

\bibitem[\protect\citeauthoryear{Magorrian et~al.,}{Magorrian
  et~al.}{1998}]{Magorrian1998}
Magorrian J.,  et~al., 1998, \mn@doi [The Astronomical Journal]
  {10.1086/300353}, 115, 2285

\bibitem[\protect\citeauthoryear{Marocco, Hache  \& Lamareille}{Marocco
  et~al.}{2011}]{Marocco2011}
Marocco J.,  Hache E.,   Lamareille F.,  2011, \mn@doi [Astronomy {\&}
  Astrophysics] {10.1051/0004-6361/201016143}, 531, A71

\bibitem[\protect\citeauthoryear{Martin et~al.,}{Martin
  et~al.}{2005}]{Martin2005}
Martin D.~C.,  et~al., 2005, \mn@doi [The Astrophysical Journal]
  {10.1086/426387}, 619, L1

\bibitem[\protect\citeauthoryear{Matsuhara et~al.,}{Matsuhara
  et~al.}{2006}]{Matsuhara2006}
Matsuhara H.,  et~al., 2006, \mn@doi [Publications of the Astronomical Society
  of Japan] {10.1093/pasj/58.4.673}, 58, 673

\bibitem[\protect\citeauthoryear{Nayyeri et~al.,}{Nayyeri
  et~al.}{2018}]{Nayyeri2018}
Nayyeri H.,  et~al., 2018, \mn@doi [The Astrophysical Journal Supplement
  Series] {10.3847/1538-4365/aaa07e}, 234, 38

\bibitem[\protect\citeauthoryear{Ng}{Ng}{2017}]{Ng2017}
Ng A.,  2017, Why is Deep Learning taking off?, \url
  {https://www.coursera.org/lecture/neural-networks-deep-learning/why-is-deep-learning-taking-off-praGm}

\bibitem[\protect\citeauthoryear{Oi et~al.,}{Oi et~al.}{2014}]{Oi2014}
Oi N.,  et~al., 2014, \mn@doi [Astronomy {\&} Astrophysics]
  {10.1051/0004-6361/201322561}, 566, A60

\bibitem[\protect\citeauthoryear{Oi, Goto, Malkan, Pearson  \& Matsuhara}{Oi
  et~al.}{2017}]{Oi2017}
Oi N.,  Goto T.,  Malkan M.,  Pearson C.,   Matsuhara H.,  2017, \mn@doi
  [Publications of the Astronomical Society of Japan] {10.1093/pasj/psx053}, 69

\bibitem[\protect\citeauthoryear{Oi et~al.,}{Oi et~al.}{2020}]{Oi2020}
Oi N.,  et~al., 2020, \mn@doi [Monthly Notices of the Royal Astronomical
  Society] {10.1093/mnras/staa3080}

\bibitem[\protect\citeauthoryear{Palanque-Delabrouille
  et~al.,}{Palanque-Delabrouille et~al.}{2011}]{PalanqueDelabrouille2011}
Palanque-Delabrouille N.,  et~al., 2011, \mn@doi [Astronomy {\&} Astrophysics]
  {10.1051/0004-6361/201016254}, 530, A122

\bibitem[\protect\citeauthoryear{Pearson et~al.,}{Pearson
  et~al.}{2017}]{Pearson2017}
Pearson C.,  et~al., 2017, \mn@doi [Publications of The Korean Astronomical
  Society] {10.5303/PKAS.2017.32.1.219}, 32, 219

\bibitem[\protect\citeauthoryear{Pearson et~al.,}{Pearson
  et~al.}{2018}]{Pearson2018}
Pearson C.,  et~al., 2018, \mn@doi [Publications of the Astronomical Society of
  Japan] {10.1093/pasj/psy107}, 71

\bibitem[\protect\citeauthoryear{Pickles}{Pickles}{1998}]{Pickles1998}
Pickles A.~J.,  1998, \mn@doi [Publications of the Astronomical Society of the
  Pacific] {10.1086/316197}, 110, 863

\bibitem[\protect\citeauthoryear{Poliszczuk et~al.,}{Poliszczuk
  et~al.}{2019}]{Poliszczuk2019}
Poliszczuk A.,  et~al., 2019, \mn@doi [Publications of the Astronomical Society
  of Japan] {10.1093/pasj/psz043}, 71

\bibitem[\protect\citeauthoryear{Richards et~al.,}{Richards
  et~al.}{2003}]{Richards2003}
Richards G.~T.,  et~al., 2003, \mn@doi [The Astronomical Journal]
  {10.1086/377014}, 126, 1131

\bibitem[\protect\citeauthoryear{Richards et~al.,}{Richards
  et~al.}{2006}]{Richards2006}
Richards G.~T.,  et~al., 2006, \mn@doi [The Astrophysical Journal Supplement
  Series] {10.1086/506525}, 166, 470

\bibitem[\protect\citeauthoryear{Ross et~al.,}{Ross et~al.}{2012}]{Ross2012}
Ross N.~P.,  et~al., 2012, \mn@doi [The Astrophysical Journal Supplement
  Series] {10.1088/0067-0049/199/1/3}, 199, 3

\bibitem[\protect\citeauthoryear{Shim et~al.,}{Shim et~al.}{2013}]{Shim2013}
Shim H.,  et~al., 2013, \mn@doi [The Astrophysical Journal Supplement Series]
  {10.1088/0067-0049/207/2/37}, 207, 37

\bibitem[\protect\citeauthoryear{Srivastava, Hinton, Krizhevsky, Sutskever  \&
  Salakhutdinov}{Srivastava et~al.}{2014}]{Srivastava2014}
Srivastava N.,  Hinton G.,  Krizhevsky A.,  Sutskever I.,   Salakhutdinov R.,
  2014, Journal of Machine Learning Research, 15, 1929

\bibitem[\protect\citeauthoryear{Stern et~al.,}{Stern et~al.}{2005}]{Stern2005}
Stern D.,  et~al., 2005, \mn@doi [The Astrophysical Journal] {10.1086/432523},
  631, 163

\bibitem[\protect\citeauthoryear{Veilleux \& Osterbrock}{Veilleux \&
  Osterbrock}{1987}]{Veilleux1987}
Veilleux S.,  Osterbrock D.~E.,  1987, \mn@doi [The Astrophysical Journal
  Supplement Series] {10.1086/191166}, 63, 295

\bibitem[\protect\citeauthoryear{Wang et~al.,}{Wang et~al.}{2020}]{Wang2020}
Wang T.-W.,  et~al., 2020, \mn@doi [Monthly Notices of the Royal Astronomical
  Society] {10.1093/mnras/staa2988}, 499, 4068

\bibitem[\protect\citeauthoryear{Webster, Francis, Petersont, Drinkwater  \&
  Masci}{Webster et~al.}{1995}]{Webster1995}
Webster R.~L.,  Francis P.~J.,  Petersont B.~A.,  Drinkwater M.~J.,   Masci
  F.~J.,  1995, \mn@doi [Nature] {10.1038/375469a0}, 375, 469

\bibitem[\protect\citeauthoryear{Yan et~al.,}{Yan et~al.}{2011}]{Yan2011}
Yan R.,  et~al., 2011, \mn@doi [The Astrophysical Journal]
  {10.1088/0004-637x/728/1/38}, 728, 38

\bibitem[\protect\citeauthoryear{Zhang \& Hao}{Zhang \& Hao}{2018}]{Zhang2018}
Zhang K.,  Hao L.,  2018, \mn@doi [The Astrophysical Journal]
  {10.3847/1538-4357/aab207}, 856, 171

\bibitem[\protect\citeauthoryear{Zhang, Schlegel, Andrews, Comparat, Schäfer,
  Mata, Kneib  \& Yan}{Zhang et~al.}{2019}]{Zhang2019}
Zhang K.,  Schlegel D.~J.,  Andrews B.~H.,  Comparat J.,  Schäfer C.,  Mata J.
  A.~V.,  Kneib J.-P.,   Yan R.,  2019, \mn@doi [The Astrophysical Journal]
  {10.3847/1538-4357/ab397e}, 883, 63

\bibitem[\protect\citeauthoryear{and}{and}{2019}]{Miyaji2019}
and T.~M.,  2019, \mn@doi [Proceedings of the International Astronomical Union]
  {10.1017/s1743921319002540}, 15, 172

\makeatother
\end{thebibliography}

\bsp	
\label{lastpage}
\end{document}